\documentclass[sigplan,screen]{acmart}

\usepackage{hyperref}
\usepackage{url}
\usepackage{xurl}
\usepackage{graphicx}
\usepackage{caption}
\usepackage{subcaption}
\usepackage{array}
\usepackage{booktabs}
\usepackage{multirow}
\usepackage{balance}
\usepackage{microtype}
\usepackage{tikz}
\usetikzlibrary{positioning, arrows.meta, shapes, calc}

\tikzset{
    >={Latex[length=3mm]}, 
    every node/.style={font=\small}, 
    every path/.style={thick} 
}

\usepackage{amsmath}
\usepackage{amssymb}

\title{Evaluating LLM-Generated ACSL Annotations for Formal Verification}

\author{Arshad Beg}
\affiliation{
  \institution{Maynooth University}
  \country{Ireland}
}
\email{arshad.beg@mu.ie}

\author{Diarmuid O'Donoghue}
\affiliation{
  \institution{Maynooth University}
  \country{Ireland}
}
\email{diarmuid.odonoghue@mu.ie}

\author{Rosemary Monahan}
\affiliation{
  \institution{Maynooth University}
  \country{Ireland}
}
\email{rosemary.monahan@mu.ie}

\begin{document}

\begin{abstract}
Formal specifications are crucial for building verifiable and dependable software systems, yet generating accurate and verifiable specifications for real-world C programs remains challenging. This paper presents an empirical evaluation of automated ACSL annotation generation strategies for C programs, comparing a rule-based Python script, Frama-C's RTE plugin, and three large language models (DeepSeek-V3.2, GPT-5.2, and OLMo 3.1 32B Instruct). The study focuses on one-shot annotation generation, assessing how these approaches perform when directly applied to verification tasks. Using a filtered subset of the CASP benchmark, we evaluate generated annotations through Frama-C's WP plugin with multiple SMT solvers, analyzing proof success rates, solver timeouts, and internal processing time. Our results show that rule-based approaches remain more reliable for verification success, while LLM-based methods exhibit more variable performance. These findings highlight both the current limitations and the potential of LLMs as complementary tools for automated specification generation.
\end{abstract}

\keywords{Static Program Analysis, ACSL Annotation, Runtime Error Detection, Specification Automation}

\maketitle

\section{Introduction}
\label{sec:introduction}

Over the past decade, artificial intelligence has advanced rapidly. Large models now perform impressively in language understanding, vision, and decision-making. Yet these systems remain difficult to trust. Their behaviour is often opaque, failures can occur silently, and there are no built-in guarantees of safety or correctness—limitations that are especially problematic in high-assurance domains. Formal methods address this gap by offering mathematically grounded guarantees about system behaviour. They allow developers to express precise requirements and prove that software satisfies them. However, formal methods are costly to apply in practice. Writing and maintaining specifications is challenging, proofs scale poorly, and purely symbolic techniques struggle with modern and evolving codebases.

This paper empirically evaluates the extent to which formal analysis tools can automatically generate and verify ACSL specifications for C programs with the help of LLMs. This work differs fundamentally from our prior study in \citep{Beg2025Traceable} in both dataset and methodology. We evaluate a new dataset, CASP released in November 2025, consisting of 506 C programs \citep{CASPdataset-AISoLA2025}. The authors of \citep{CASPdataset-AISoLA2025} has used the dataset with Gemini-based prompt feedback loops to iteratively correct programs or annotations reported in \citep{VeCoGen-Formalise-ICSE2025}. Here, we reuse the CASP dataset as a basis for evaluation across our annotation strategies. Here, we conduct a controlled comparison of five ACSL generation strategies: a rule-based Python script, Frama-C's RTE plugin, and three LLMs--DeepSeek-V3.2, GPT-5.2, and OLMo 3.1 32B Instruct. In the paper, we refer to LLMs as DeepSeek, GPT-5, and OLMo3, respectively, without using their full names. This differs from prior work that focuses either on dataset construction \cite{CASPdataset-AISoLA2025} or on individual specification generation technique in isolation \cite{VeCoGen-Formalise-ICSE2025}. All generated specifications are verified under identical conditions using the Frama-C WP plugin and multiple SMT solvers. This design enables a direct comparison between tool generated and LLM generated ACSL on a modern dataset, isolating the impact of annotation quality, solver sensitivity, and proof stability, and providing new empirical evidence that complements, but does not overlap with our earlier work \cite{BegEtAl2025, Beg2025Traceable}, as we used different dataset of C Programs and workflow.

\paragraph{Contributions.}
The contribution of paper is an empirical comparison of five ACSL annotation generation strategies, including rule-based, tool-supported, and LLM-based approaches, under same settings. It evaluates the effectiveness of generated annotations using Frama-C's WP plugin across multiple SMT solvers, reporting proof success rates, timeouts, and processing metrics. It further analyzes the practical strengths and limitations of each approach, highlighting differences in reliability and variability across strategies.

\paragraph{Terminology for clarification.}
We briefly clarify key terms used in the evaluation. \textit{Qed} refers to the internal simplification phase of the Frama-C WP plugin, which attempts to discharge proof obligations through built-in reasoning before invoking external solvers. \textit{Kernel warnings} denote messages generated by the Frama-C analysis kernel when encountering unsupported constructs, potential inconsistencies, or issues during code processing.

The structure of the paper is as follows: Most recent literature and its analysis is presented in Section \ref{sec:related_work}. Section \ref{sec:experimental_setup} is about the experimental setup and the workflow we adopted to conduct the research. Empirical evaluation of our experiments is presented in Section \ref{sec:empirical_evaluation} and Section \ref{sec:conclusion} concludes the paper. 

\section{Related Work}
\label{sec:related_work}

Our earlier workshop paper \cite{BegEtAl2025} examined the challenges of translating informal natural-language requirements into formal specifications, highlighting issues such as ambiguity, missing domain knowledge, contextual gaps, and instability in large language model outputs. That work introduced the VERIFYAI framework, which integrates LLMs with NLP pipelines, ontology-based modelling, and artefact reuse to support semi-automated formalisation. It also compared systems such as Req2Spec \cite{Req2SpecPaper}, SpecGen \cite{Ma2024}, AssertLLM \cite{10691792}, and nl2spec \cite{cosler2023nl2specinteractivelytranslatingunstructured}, analysing their approaches to requirement representation and verification integration. An empirical evaluation using Frama-C PathCrawler and SMT solvers exposed limitations in verification scalability and solver robustness.

These ideas were substantially extended in \citep{Beg2025Traceable}, a large-scale survey of over one hundred studies on AI-enabled requirements formalisation, traceability, and verification. The survey mapped the broader research landscape across programming languages, specification notations, and verification backends, including both theorem provers and model checkers. It identified recurring methodological patterns, from rule-based extraction to hybrid neuro-symbolic pipelines, and highlighted unresolved ecosystem challenges such as weak toolchain integration, limited traceability, and poor reuse of existing specification artefacts. In addition, the study reported an expanded empirical evaluation of Frama-C's EVA, RTE, and PathCrawler tools, revealing recurring issues related to solver instability, path coverage, and configuration sensitivity. Together, these prior works provide the methodological and empirical foundation for the focused analysis presented in this paper.

PALM, introduced in \citep{PALMforCoqProofs2024}, is a generate-and-repair framework that integrates large language models with symbolic reasoning to enhance formal proof generation in Coq. Analyzing 520 proof-generation errors by GPT-3.5, Minghai et al. observed that while LLMs often capture the overarching structure of proofs, they frequently fail at low-level steps. This insight motivated PALM's iterative repair mechanism. Tested on a dataset exceeding 10,000 theorems, PALM demonstrates a 76.6\%-180.4\% improvement in success rates, proving an additional 1,270 theorems over previous methods and showing strong generalizability across multiple LLM architectures.

The framework AutoSpec, presented in \citep{AutoSpecCAV2024}, automates the synthesis of formal specifications to enable end-to-end program verification with minimal manual intervention. Unlike earlier approaches constrained to narrow domains, AutoSpec handles complex program constructs, including arrays, pointers, nested loops, and function calls. The framework employs an iterative synthesis loop that combines static analysis and program verification, progressively refining candidate specifications while ensuring both satisfiability and proof adequacy. Empirical evaluation indicates robust performance, successfully verifying 79\% of benchmark programs, a 1.592 times improvement over prior techniques and further validating its practicality by verifying the real-world X509-parser project.

In \citep{fan2025evaluatingabilitylargelanguage}, the authors assessed GPT-4o’s capability to generate verifiable specifications for C programs using VeriFast, a separation-logic-based static verifier. Their experiments, which varied user inputs and prompting strategies, revealed that GPT-4o could preserve functional behavior in specifications; however, the generated specifications often failed formal verification and contained redundancies. Similarly, \citep{LLMs4OREZhao202503} introduces OntoChat, a conversational agent leveraging LLMs to support the Ontology Requirements Engineering (ORE) process. OntoChat addresses limitations of manual ORE methods by facilitating automated and enhanced requirements elicitation, documentation, and validation. Preliminary findings from the first year indicate that LLM assisted interactions can improve efficiency, consistency, and collaboration in ontology engineering.

HILBERT, detailed in \citep{HilbertArXiV202509}, is an integrated agentic framework combining informal mathematical reasoning with formal verification to advance automated theorem proving. The system coordinates four synergistic components: an informal reasoning model for high-level insights of Lean 4, which is an optimised prover, a formal verifier, and a semantic theorem retriever. For unresolved problems, HILBERT decomposes them recursively into smaller subgoals, addressed via either the prover or the informal reasoning agent. Verifier-guided feedback loops iteratively correct and reinforce proofs, maintaining semantic consistency. Experimental results demonstrate state-of-the-art performance: 99.2\% accuracy on miniF2F, 6.6 percentage points higher than prior best methods and 462/660 problems (70.0\%) solved on PutnamBench, surpassing proprietary systems like SeedProver (50.4\%) and improving 422\% over the best publicly available baseline. These outcomes establish HILBERT as a significant advancement in bridging informal reasoning and formally verified proof synthesis.

LEMUR, introduced in \citep{LemurWu2024}, is a framework that combines automated reasoning with structured synthesis to improve loop invariant generation and program verification. Unlike prior learning-based methods such as Code2Inv, which rely purely on reinforcement learning, LEMUR formalizes a unified calculus integrating symbolic reasoning and guided invariant inference. On the Code2Inv benchmark set of 133 C programs, LEMUR uses the ESBMC k-induction verifier to validate invariants, significantly outperforming ESBMC alone and Code2Inv. Specifically, LEMUR (GPT-4) solves 107 benchmarks within a 10-minute timeout, compared to 68 for ESBMC and fewer for Code2Inv under a one-hour limit. Its adaptive generation strategy typically converges in four iterations, though complex cases may take longer. Comparative results with LEMUR (GPT-3.5) highlight that stronger symbolic reasoning oracles lead to faster convergence and higher verification success, emphasizing the value of structured reasoning over heuristic generation.

RvLLM, described in \citep{RvLLM202505}, provides a runtime verification framework that exploits domain-specific knowledge encoded in a custom specification language (ESL) to systematically detect and correct errors. Operating in two stages, interpretation and reasoning, the framework uses context-driven interpretations to identify inconsistencies and iterative follow-up queries to ensure output consistency. Evaluations across three representative tasks—violation detection under the Singapore Rapid Transit Systems Act, numerical comparison, and inequality solving—tested multiple LLMs including Qwen (max, plus, turbo, 2.5 variants), GPT-4.1 (mini, nano), Gemini 2.0 Flash (Lite), and DeepSeek-V3. In violation detection, RvLLM substantially improved true positive rates, raising Qwen max from 56.2\% to 86.1\% and GPT-4.1 from 57.7\% to 81.1\%, while maintaining true negative rates. In numerical comparison, LLMs guided by RvLLM consistently produced correct or inconclusive results, e.g., Qwen 2.5 (32B) yielded 98 correct and 2 inconclusive outputs in 4.98 seconds. In inequality solving tasks involving factorization, interval analysis, and endpoint checking, RvLLM improved true positive rates, achieving 50\% for Qwen 2.5 on factorization, demonstrating effective enforcement of domain-specific constraints.

The most recent works of November 2025 have following contribution: \citep{VeCoGen-Formalise-ICSE2025} introduces VeCoGen, a tool that combines large language models with formal verification to automatically generate correct C programs from formal and natural language specifications along with test cases. VeCoGen generates candidate programs and iteratively refines them until a program satisfies the formal specification, ensuring correctness. The approach is evaluated on 15 Codeforces problems, successfully solving 13, demonstrating the feasibility of integrating LLMs with formal methods for program generation. This work highlights the potential of leveraging LLMs to produce formally verified code suitable for safety-critical applications. \citep{CASPdataset-AISoLA2025} addresses the lack of datasets for benchmarking verified C code generation by presenting CASP, a curated dataset of 506 C programs paired with formally verified ACSL specifications. The dataset is constructed through multi-stage filtering, formal verification using Frama-C, LLM-assisted improvements, and manual inspection to ensure correctness. CASP enables systematic evaluation of automated code generation methods against verified specifications. This contribution provides a foundation for research on integrating LLM-based code generation with formal verification.

SpecVerify, presented in \citep{MarieFarrell202507}, integrates LLMs with formal verification tools to automatically extract and verify properties from natural language requirements. By combining Claude 3.5 Sonnet with the ESBMC verifier, SpecVerify achieves verification accuracy comparable to NASA's CoCoSim, while reducing false positives and extending the expressiveness of assertions beyond traditional logics. Complementing this, \citep{FindingLoopInvariantsArxiv2023} explores methods for automatically discovering inductive loop invariants, a long-standing formal verification challenge. Using a benchmark of 1,025 C programs with diverse loops, GPT-4, GPT-3.5, and Code Llama were paired with Frama-C's WP tool and SMT solvers to evaluate invariant soundness. The approach synthesizes candidate invariants through data-driven methods and validates them formally via automated proof obligations.

SV-LLM, introduced in \citep{SVLLM4SoCDesign202506}, is a multi-agent LLM framework designed for automating security verification of complex system-on-chip (SoC) designs. The framework deploys domain-specialized agents to perform tasks such as asset identification, threat modeling, property generation, vulnerability detection, and simulation-based bug validation. Agents employ diverse learning paradigms—including fine-tuning, in-context learning, and retrieval-augmented generation—to produce precise, verifiable security specifications from design data and documentation, with reasoning capabilities tailored to RTL semantics and design constraints. Experimental evaluation shows substantial gains: the fine-tuned Security Vulnerability Detection Agent (Mistral-7B-Instruct) achieved 84.8\% accuracy, a 42.3 percentage-point improvement over its non-fine-tuned baseline, while the Bug Validation Agent reached up to 89\% validated testbench generation, outperforming zero-shot prompting by more than fourfold. These results demonstrate SV-LLM as a scalable, explainable, and domain-adapted solution for SoC security verification. Finally, \citep{Granberry2025} explores integrating Copilot with formal methods through an IDE equipped with language servers. Granberry et al. \citep{Granberry2025a} also studied combining LLMs with symbolic analysis for C program specification generation, enhancing LLM prompts using outputs from PathCrawler and EVA to produce ACSL annotations.

\paragraph{Positioning of this work.}
While prior research has explored specification generation using both rule-based systems and learning-based approaches, these efforts are typically evaluated in isolation or under differing experimental conditions. In contrast, this paper provides a unified empirical comparison of multiple annotation generation strategies within a consistent verification pipeline, enabling a direct assessment of their practical effectiveness.

\section{Experimental Setup and Workflow}
\label{sec:experimental_setup}

We evaluate five approaches for generating ACSL annotations. A rule-based Python script inserts annotations from predefined patterns targeting safety properties such as null checks, array bounds, and basic preconditions, without inferring functional specifications. The Frama-C RTE plugin generates annotations to prevent runtime errors (e.g., division by zero, invalid memory access, overflows) but is limited to safety properties. We also evaluate three LLMs (DeepSeek, GPT-5, and OLMo3), each prompted in a one-shot setting to produce syntactically valid ACSL annotations, including \texttt{requires}, \texttt{ensures}, and loop invariants, without fine-tuning or iterative feedback. All LLMs use a consistent prompt template and fixed hyper-parameters. The approaches differ in scope: RTE focuses on runtime safety, the rule-based method covers a subset of safety annotations, and LLMs may generate both safety and functional specifications.

\paragraph{Evaluation rationale.}
The inclusion of rule-based and tool-supported approaches (e.g., the Python script and Frama-C RTE plugin) is motivated by our prior work \cite{BegEtAl2025,Beg2025Traceable}, where these techniques were systematically studied for ACSL annotation generation and verification. As the dataset used in this paper differs entirely from those earlier studies, we first re-evaluate these established approaches on the new dataset to provide a consistent baseline. This enables a controlled comparison before extending the evaluation to LLM-based methods, ensuring that observed differences are attributable to the generation strategies rather than dataset variation.

\begin{figure*}[t]
\centering
\resizebox{\textwidth}{!}{%
\begin{tikzpicture}[every node/.style={draw, rounded corners, align=center, minimum width=2.5cm, minimum height=0.6cm,fill=blue!20}, 
    >=Latex, 
    scale=1,
    transform shape]

\node (curated) at (0,8.0) {\small{Curated Dataset}};
\node (pure) at (2.0,7.0) {\small{Pure C files}};
\node (acsl) at (4.0,6.0) {\small{ACSL Annotations}};
\node (annotated) at (6.0,5.0) {\small{Annotated C files}};
\node (wp) at (8.0,4.0) {\small{WP Provers Tests}};
\node (results) at (10.0,3.0) {\small{Results}};
\node (analysis) at (12.0,2.0) {\small{Analysis}};

\draw[->, thick] (curated) -- (pure);
\draw[->, thick] (pure) -- (acsl);
\draw[->, thick] (acsl) -- (annotated);
\draw[->, thick] (annotated) -- (wp);
\draw[->, thick] (wp) -- (results);
\draw[->, thick] (results) -- (analysis);

\node (t1) at (8.5,7.5) {\small{Python Script, Frama-C RTE, DeepSeek, GPT-5, OLMo-3}};

\draw[->, thick] (t1.south) -- (acsl.north);

\end{tikzpicture}%
}
\caption{End-to-end research workflow illustrating how a curated dataset is progressively transformed into pure C files, enriched with ACSL annotations generated via a combination of automated scripts, Frama-C RTE, and large language models (DeepSeek-V3.2, GPT-5.2, and OLMo-3.1 32B Instruct), then consolidated into annotated C files that are evaluated using weakest-precondition (WP) prover tests to produce verification results that are finally subjected to systematic analysis.}
\label{fig:research_workflow}
\end{figure*}
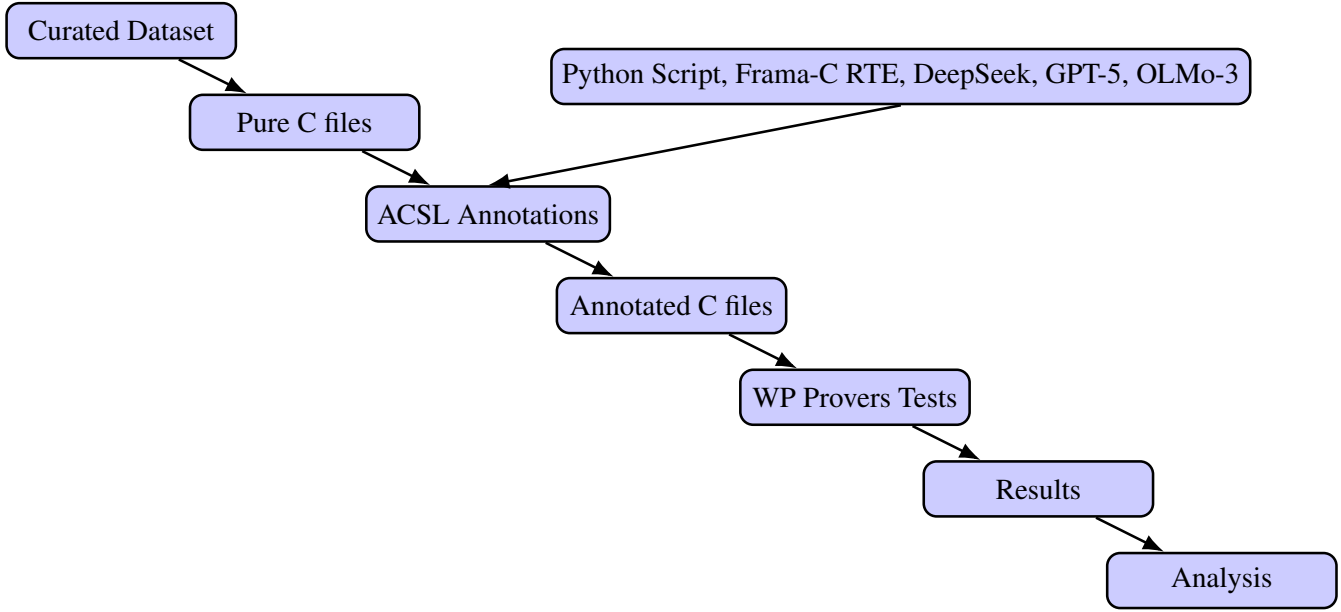

All artifacts are available at \url{https://github.com/arshadbeg/FTfJPatECOOP2026}. Experiments run in a controlled WSL environment on Windows 11 (Intel Core Ultra 5 125U, 32 GB RAM, 64-bit) to ensure toolchain compatibility. The dataset from \url{https://huggingface.co/datasets/nicher92/CASP_source_files}~\citep{CASPdataset-AISoLA2025} is preprocessed into standard-compliant C files. Of 506 files, 355 valid programs are retained after excluding 151 empty or incomplete ones. The workflow (Figure~\ref{fig:research_workflow}) covers dataset preparation, annotation generation, weakest-precondition verification, and results analysis, with all LLMs evaluated using default decoding settings (e.g., temperature 1.0).

\section{Empirical Evaluation}
\label{sec:empirical_evaluation}

\subsection{EVA Static Analysis}

\begin{figure*}[h]
    \centering
    \begin{subfigure}[t]{0.32\textwidth}
        \centering
        \includegraphics[width=\textwidth]{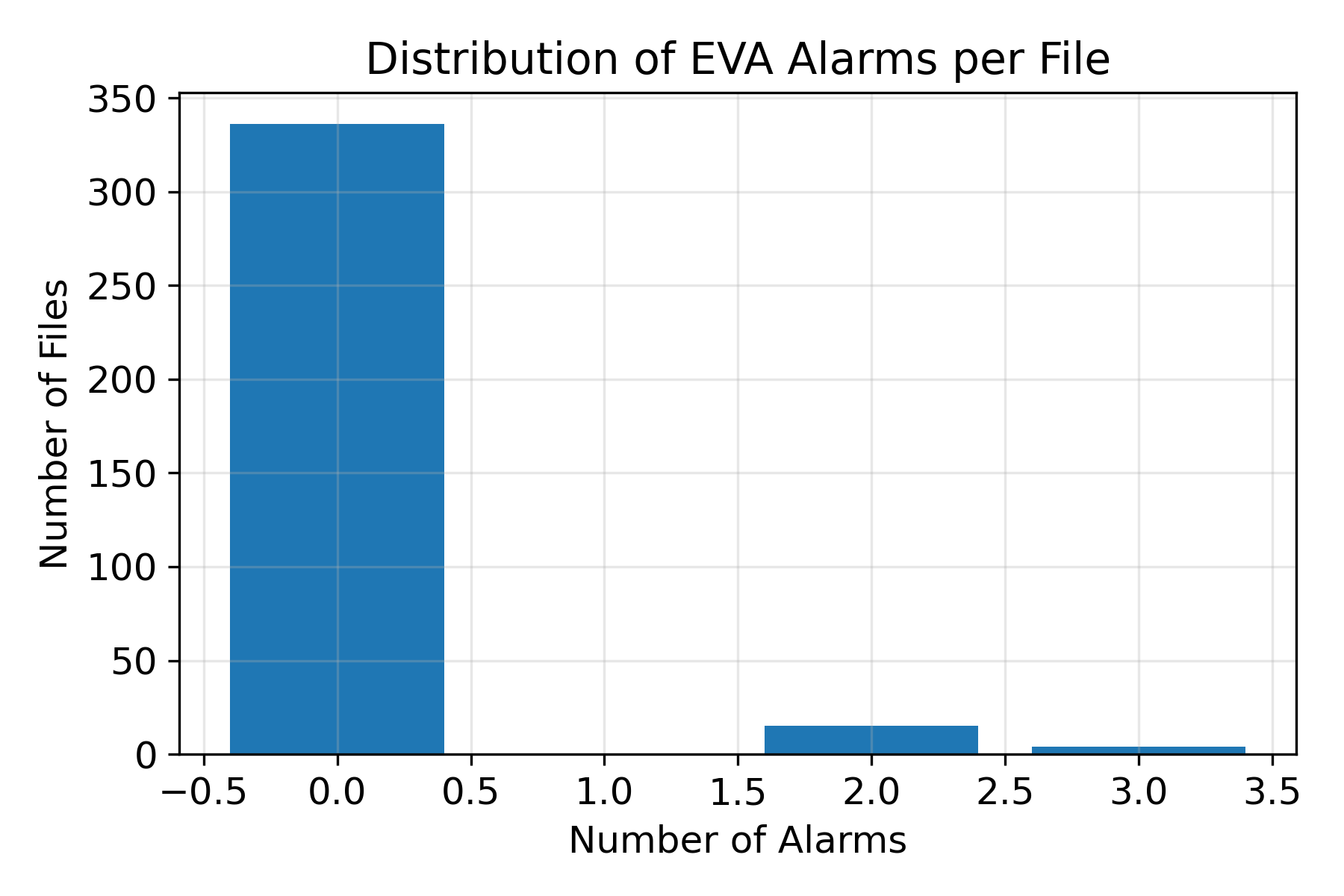}
        \caption{Distribution of EVA alarms per file.}
        \label{fig:alarms_hist}
    \end{subfigure}
    \hfill
    \begin{subfigure}[t]{0.32\textwidth}
        \centering
        \includegraphics[width=\textwidth]{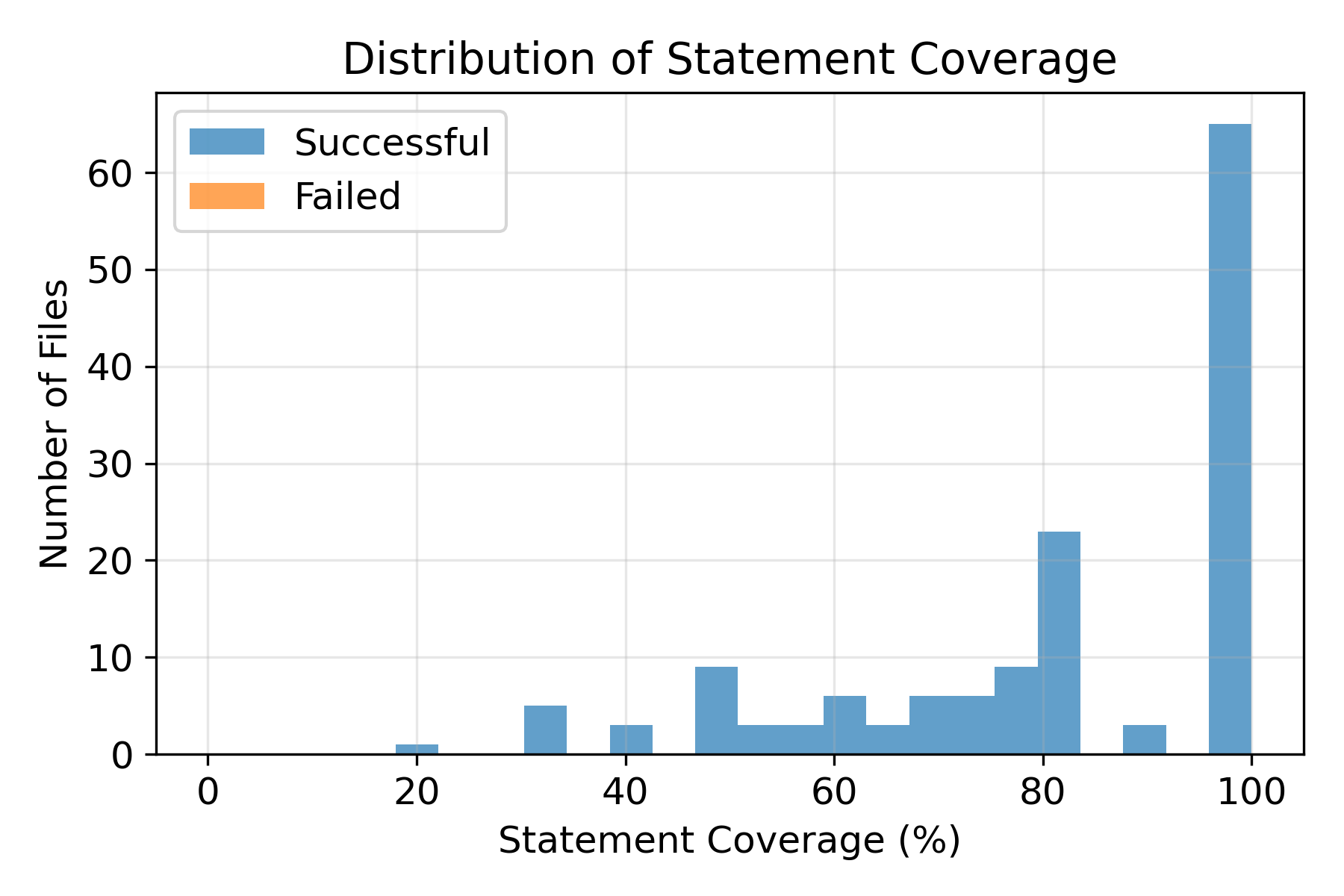}
        \caption{Histogram of statement coverage for successful and failed analyses.}
        \label{fig:coverage_hist}
    \end{subfigure}
    \hfill
    \begin{subfigure}[t]{0.32\textwidth}
        \centering
        \includegraphics[width=\textwidth]{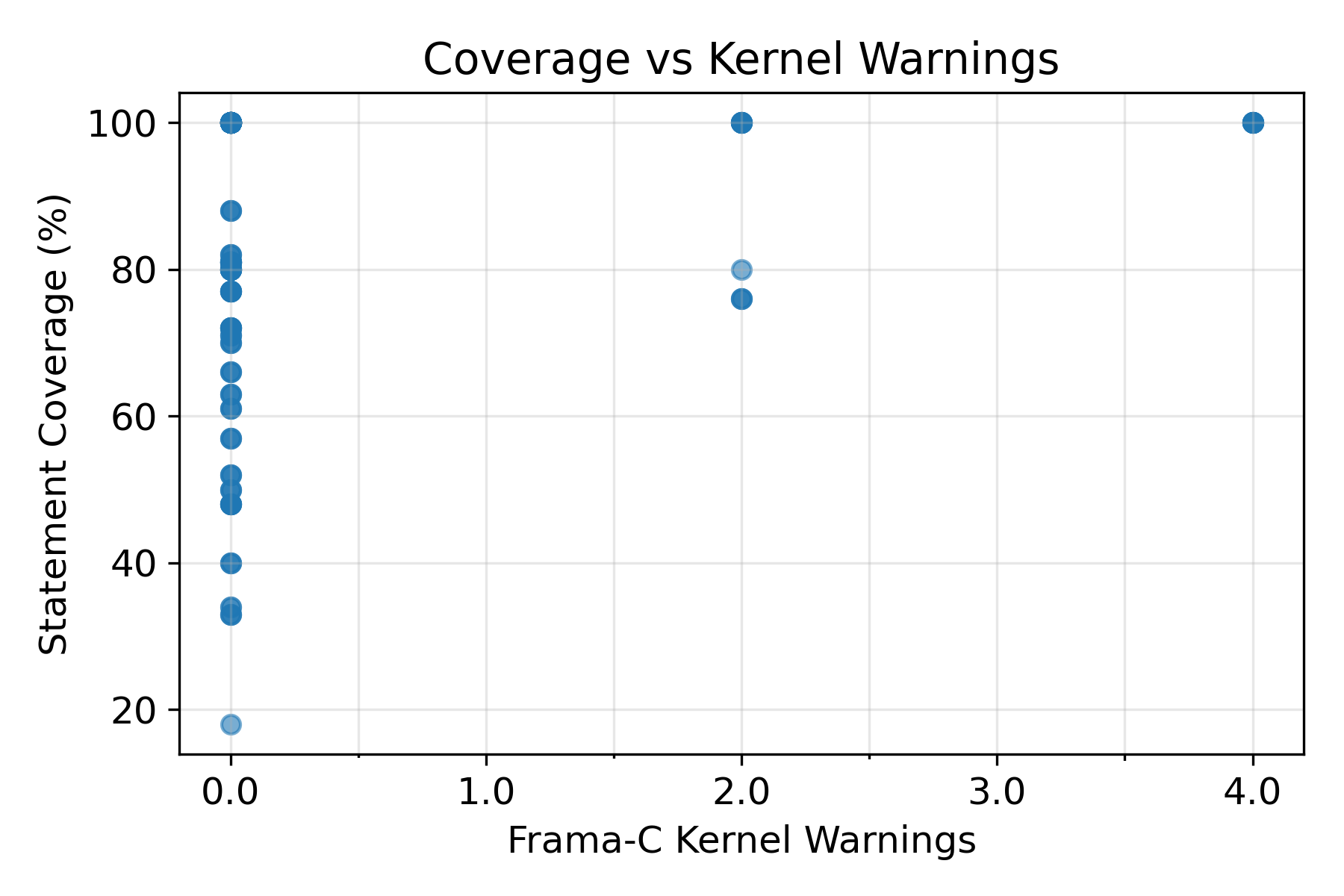}
        \caption{Statement coverage vs Frama-C kernel warnings.}
        \label{fig:coverage_vs_warnings}
    \end{subfigure}
    \caption{Visual summary of EVA analysis results. (a) Number of alarms per file. (b) Distribution of statement coverage for successful and failed analyses. (c) Relationship between kernel warnings and coverage.}
    \label{fig:eva_analysis_summary}
\end{figure*}

We first performed a static analysis on these pure C files through Frama-C EVA tool. Table \ref{tab:eva_summary} provides a summary of results from EVA, augmented with the figure \ref{fig:eva_analysis_summary}. Although EVA successfully analysed only 40.8\% of the files, we achieved a median statement coverage of 81\%, with alarm generation remaining sparse across the dataset. 

\begin{table}[h!]
\centering
\caption{EVA Dataset Summary}
\resizebox{\columnwidth}{!}{%
\begin{tabular}{ccccccc}
\hline
\textbf{Dataset} & \textbf{Files} & \textbf{Mean } & \textbf{Median} & \textbf{Mean} & \textbf{Max} & \textbf{Kernel} \\
 &  & \textbf{coverage (\%)} & \textbf{coverage (\%)} & \textbf{alarms} & \textbf{alarms} & \textbf{warnings} \\
\hline
All files      & 355 & 81.41 & 81.0 & 0.12 & 3 & 19 \\
Successful     & 145 & 81.41 & 81.0 & 0.29 & 3 & 19 \\
Failed         & 210 & --    & --   & 0.00 & 0 & 0 \\
\hline
\end{tabular}
}
\label{tab:eva_summary}
\end{table}

Figure~\ref{fig:eva_analysis_summary} provides a visual overview of the EVA analysis results. Figure~\ref{fig:alarms_hist} shows the distribution of alarms per file, highlighting that most files generate few or no alarms. Figure~\ref{fig:coverage_hist} compares statement coverage between successful and failed analyses, indicating that failed analyses do not have coverage data. Figure~\ref{fig:coverage_vs_warnings} examines the relationship between kernel warnings and statement coverage, illustrating that files with more warnings tend to have lower coverage.

\paragraph{Annotation quality.}
Beyond verification success rates, the quality of generated annotations varies across approaches. In particular, some annotations—especially those produced by LLM-based methods—are syntactically valid but semantically weak, such as trivially true conditions or overly conservative constraints. While these annotations may not introduce verification failures, they do not meaningfully contribute to proving functional correctness. In contrast, rule-based approaches and the RTE plugin tend to produce more targeted annotations aligned with specific safety properties, resulting in more consistent verification outcomes. These differences highlight that verification success alone does not fully capture annotation usefulness.

\subsection{Analysis based on Mean Values}

\begin{figure*}[t]
\centering

\begin{subfigure}[t]{0.45\textwidth}
  \centering
  \includegraphics[width=\linewidth,keepaspectratio]{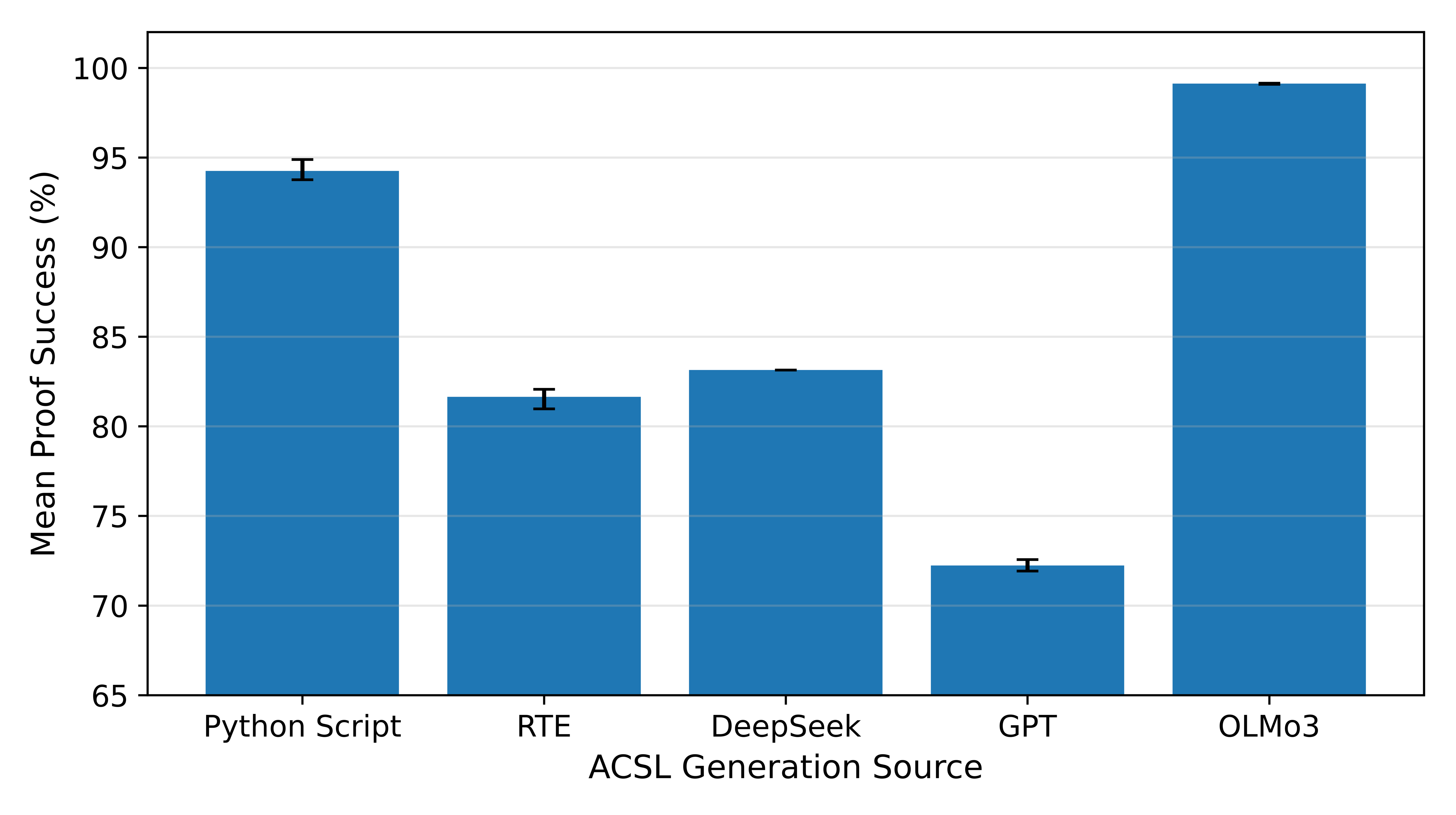}
  \caption{Mean proof success rate with solver-dependent variability.}
  \label{fig:wp_success_variability}
\end{subfigure}
\hfill
\begin{subfigure}[t]{0.45\textwidth}
  \centering
  \includegraphics[width=\linewidth,keepaspectratio]{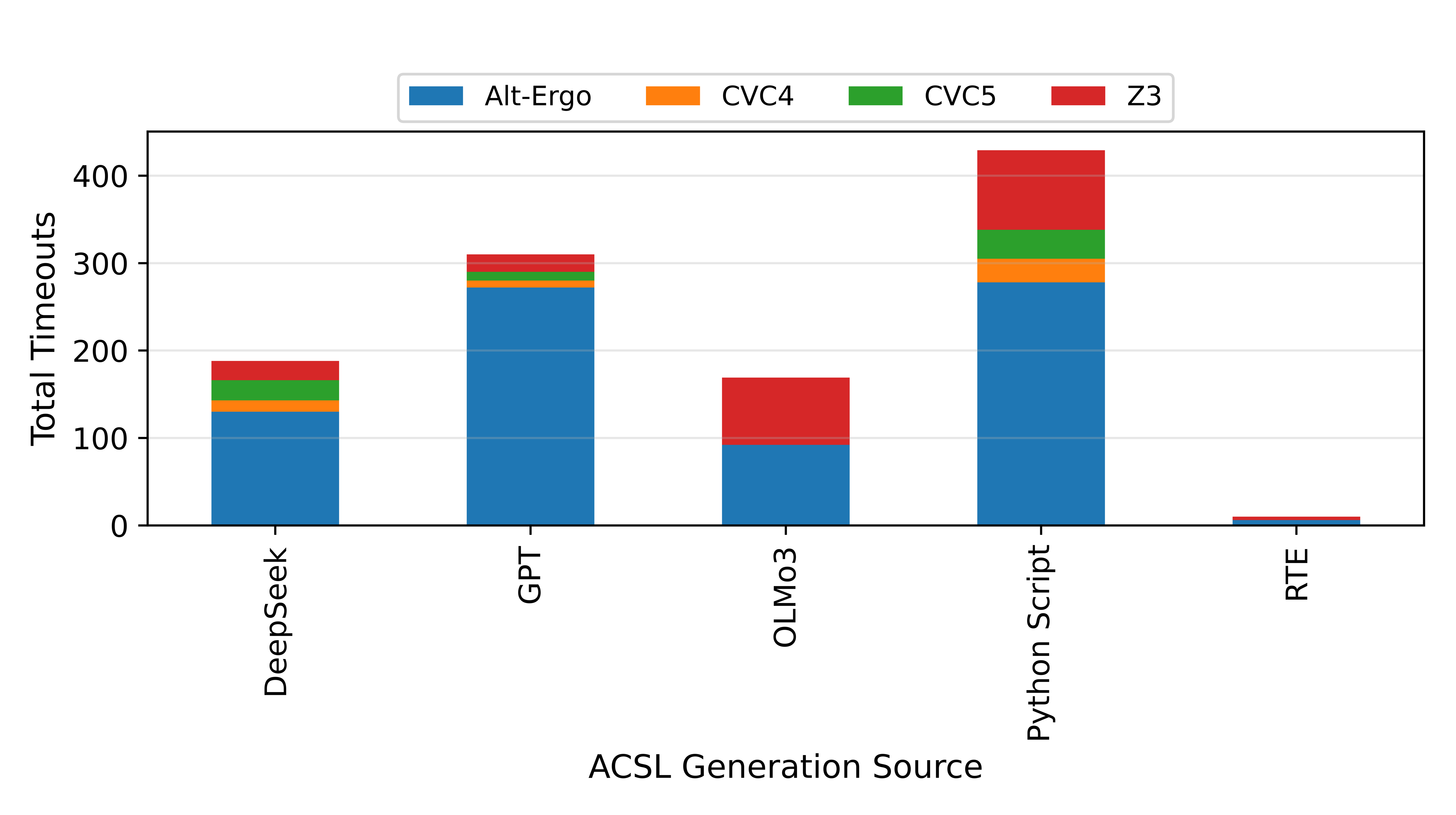}
  \caption{Distribution of solver timeouts per ACSL source.}
  \label{fig:wp_timeouts}
\end{subfigure}
\hfill
\begin{subfigure}[t]{0.7\textwidth}
  \centering
  \includegraphics[width=\linewidth,keepaspectratio]{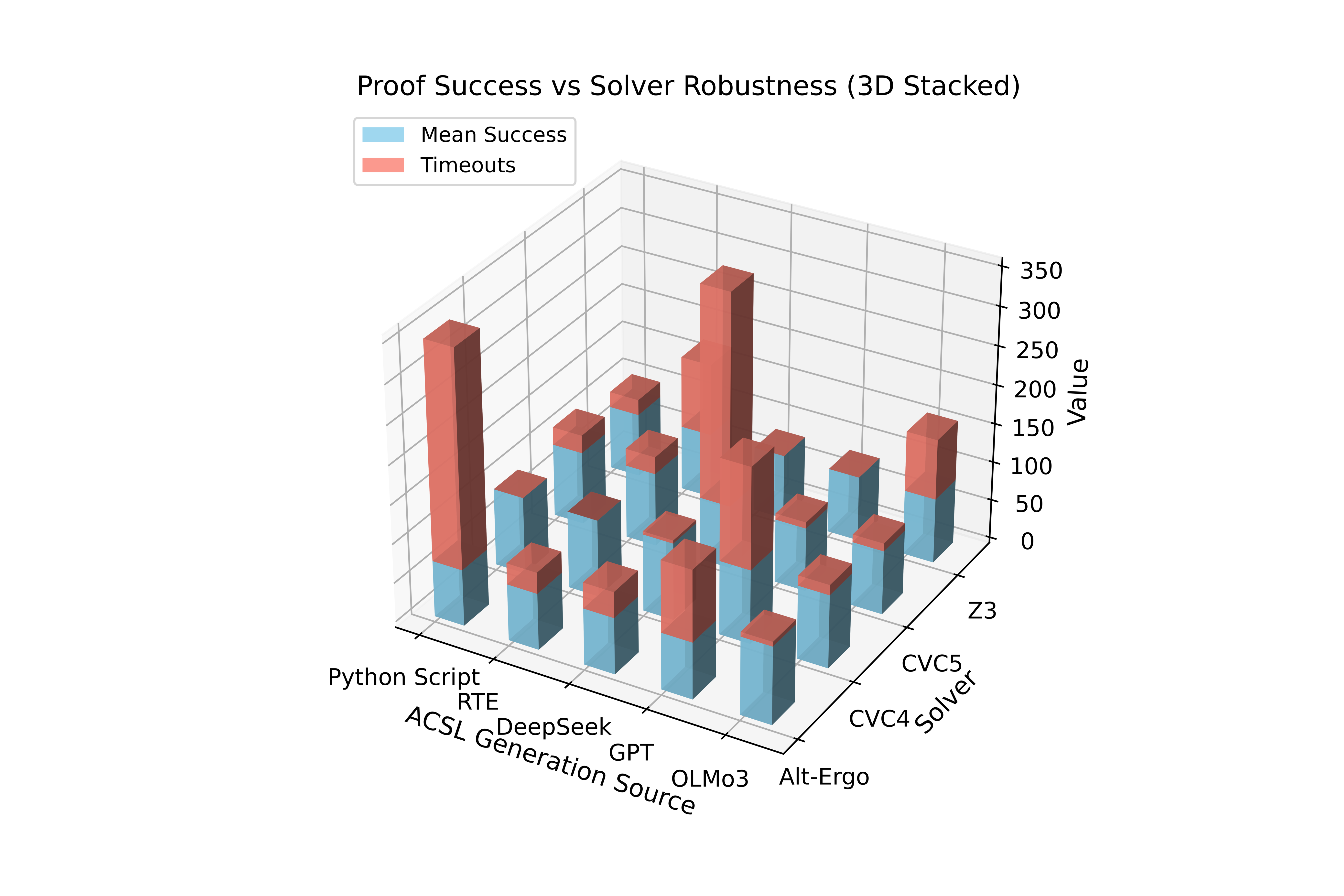}
  \caption{Trade-off between proof success and solver robustness.}
  \label{fig:wp_success_vs_timeouts}
\end{subfigure}

\caption{Comparison of WP verification outcomes across ACSL generation methods.
The figure contrasts proof success, solver robustness, and their interaction
for tool-generated and LLM-generated specifications.}
\label{fig:joint_wp_comparison}

\end{figure*}

Table~\ref{tab:wp_acsl_comparison} summarises the aggregate WP verification statistics across all ACSL generation strategies, providing a quantitative complement to the visual trends in Fig.~\ref{fig:joint_wp_comparison}. Tool-generated specifications, particularly those produced by the Python script and the RTE plugin, achieve the highest mean proof success rates while maintaining low timeout counts, confirming their reliability for automated verification. Among LLM-based approaches, DeepSeek exhibits the strongest overall performance, with mean success and timeout figures approaching those of RTE, whereas GPT and OLMo3 show reduced success rates and substantially higher solver timeouts, especially for Alt-Ergo. Notably, the median proof success reaches 100\% for all approaches, indicating that when proofs succeed, they often do so completely; however, the variability captured by the mean values and timeout counts highlights significant differences in specification robustness. Overall, Table~\ref{tab:wp_acsl_comparison} reinforces that ACSL generation quality must be evaluated not only by proof success but also by solver stability and efficiency.

\begin{table}[t]
\centering
\caption{WP verification results for ACSL annotations generated by different sources.
Mean success denotes the average percentage of proved goals over generated goals.
Qed time is reported in milliseconds.}
\label{tab:wp_acsl_comparison}

\scriptsize
\renewcommand{\arraystretch}{1.1}
\setlength{\tabcolsep}{3pt}

\resizebox{0.95\columnwidth}{!}{%
\begin{tabular}{llcccc}
\toprule
Source & Solver & Runs & Mean & TO & Qed \\
\midrule
\multirow{4}{*}{Python Script}
 & AltErgo & 207 & 71.93 & 278 & 1.25 \\
 & CVC4 & 205 & 72.23 & 27  & 1.16 \\
 & CVC5 & 203 & 72.19 & 33  & 1.16 \\
 & Z3 & 205 & 72.57 & 91  & 1.22 \\
\midrule
\multirow{4}{*}{RTE}
 & AltErgo & 115 & 99.15 & 6 & 7.50 \\
 & CVC4 & 113 & 99.12 & 0 & 8.50 \\
 & CVC5 & 111 & 99.09 & 0 & 7.50 \\
 & Z3 & 113 & 99.12 & 4 & 6.50 \\
\midrule
\multirow{4}{*}{DeepSeek}
 & AltErgo & 149 & 94.89 & 130 & 6.17 \\
 & CVC4 & 147 & 93.76 & 13  & 6.53 \\
 & CVC5 & 145 & 93.80 & 23  & 6.32 \\
 & Z3 & 147 & 94.56 & 22  & 5.04 \\
\midrule
\multirow{4}{*}{GPT}
 & AltErgo & 177 & 80.98 & 272 & 1.86 \\
 & CVC4 & 175 & 81.63 & 8   & 2.00 \\
 & CVC5 & 173 & 81.90 & 10  & 2.30 \\
 & Z3 & 175 & 82.07 & 20  & 1.97 \\
\midrule
\multirow{4}{*}{OLMo3}
 & AltErgo & 134 & 83.14 & 92 & 2.92 \\
 & CVC4 & 133 & 83.14 & 0  & 2.81 \\
 & CVC5 & 132 & 83.14 & 0  & 3.05 \\
 & Z3 & 133 & 83.14 & 77 & 2.87 \\
\bottomrule
\end{tabular}%
}

\end{table}

Figure \ref{fig:joint_wp_comparison} presents a consolidated comparison of Frama-C WP verification performance across ACSL specifications generated by the Python script, Frama-C RTE, and the three LLMs (DeepSeek, GPT, and OLMo3). As shown in Fig. \ref{fig:wp_success_variability}, tool-generated specifications (Python script and RTE) consistently achieve higher mean proof success rates with lower solver-dependent variability, reflecting more precise and verification friendly contracts. Among LLM-based approaches, DeepSeek demonstrates relatively strong performance with proof success close to tool-generated ACSL, whereas GPT and OLMo3 exhibit noticeably lower mean success and higher dispersion across solvers. This trend indicates that while LLMs can often produce syntactically valid ACSL, the semantic strength and completeness of the generated contracts vary substantially, directly impacting automated proof outcomes.

Solver robustness is further highlighted in Fig. \ref{fig:wp_timeouts}, which reports the total number of solver timeouts aggregated across all files. RTE-generated ACSL yields the fewest timeouts overall, confirming the conservative yet stable nature of runtime-error–driven specifications. In contrast, GPT-generated annotations result in the highest number of timeouts, particularly for Alt-Ergo, suggesting that overly weak or ambiguous specifications increase solver search space and proof difficulty. This trade-off is made explicit in Fig. \ref{fig:wp_success_vs_timeouts}, where a clear inverse relationship emerges between proof success and timeout frequency across ACSL sources. The X-axis of the figure represents the different ACSL generation sources used to produce verification tasks, while the Y-axis denotes the SMT solvers applied to attempt proofs on these tasks. The Z-axis shows the numerical values, with the lower portion of each bar indicating the mean proof success in percentage and the stacked portion on top representing the total number of timeouts. This arrangement allows a clear comparison of both solver effectiveness and robustness across different sources. Together, these results demonstrate that higher apparent proof success does not necessarily imply solver efficiency, and that specification quality must be assessed jointly through success rates and solver behavior, as summarized holistically in Fig. \ref{fig:joint_wp_comparison}.

\subsection{Rule-Based Python Script Generation}
\label{subsec:python_script_generation}

The approach employs a deterministic, rule-based pipeline implemented in Python to automatically augment C programs with ACSL (ANSI/ISO C Specification Language) annotations. The rules are derived from syntactic pattern matching over control-flow constructs and function structures, rather than relying on semantic program analysis. At the function level, each function body (excluding \texttt{main}) is analyzed for conditional return patterns. When an \texttt{if}-statement with corresponding \texttt{return} statements in both branches is detected, the system generates postconditions of the form:
\[
\texttt{ensures } (C) \Rightarrow \backslash result = E_1
\]
\[
\texttt{ensures } \neg(C) \Rightarrow \backslash result = E_2
\]
where $C$ denotes the normalized branch condition and $E_1$, $E_2$ are the returned expressions. This rule captures functional behavior under the assumption of a single conditional decision governing the return value.

For loop constructs (\texttt{for}, \texttt{while}, and \texttt{do}), the system applies heuristic-based rules to infer loop indices, bounds, and array accesses. Precondition clauses are generated to ensure valid memory access (e.g., \texttt{\backslash valid} over array segments) and positive bounds. Loop invariants are introduced to constrain index ranges and express properties over accessed arrays using quantified expressions. Additionally, loop termination is enforced via a variant decreasing with each iteration. Switch-case statements are handled by associating each \texttt{case} label with its corresponding return statements, producing conditional postconditions that relate case conditions to returned values. This encodes the control-flow semantics of multi-branch selection in a declarative form. Finally, at the program entry point (\texttt{main}), the system inserts runtime assertions following function calls whose results are assigned to variables. These assertions are derived from simple syntactic properties of arguments (e.g., literals or constants) and serve as lightweight checks.

Overall, the approach consists of a set of syntactic transformation rules that map recognizable code patterns to ACSL specifications, enabling automated annotation generation for structurally conventional C programs without requiring deep semantic reasoning.

Table~\ref{tab:wp-summary} summarises the results of 452 WP runs distributed across four provers, with each solver executed between 111 and 115 times. For all solvers, the observed proof success values lie between 75\% and 100\%, indicating that, in every run, at least three quarters of the generated proof obligations were discharged automatically. The dispersion of these values is limited: Alt-Ergo exhibits a standard deviation of 4.56 over 115 runs, CVC4 and Z3 both show a standard deviation of 4.64 over 113 runs, and CVC5 reaches the highest observed variability with a standard deviation of 4.72 over 111 runs. Despite minor differences, the narrow spread and identical minimum and maximum values across solvers suggest that all four provers achieve comparable proof coverage on the evaluated benchmarks, with no solver consistently failing or excelling across the full set of verification tasks.

Differences between solvers become more apparent when considering timeout behavior and proof discharge times, as reported in Table~\ref{tab:wp-summary}. Alt-Ergo incurred 6 timeouts over 115 runs, while Z3 experienced 4 timeouts over 113 runs; in contrast, both CVC4 and CVC5 completed all runs without any observed timeouts. Mean Qed times range from 6.50ms for Z3 to 8.50ms for CVC4, with Alt-Ergo and CVC5 both averaging 7.50ms. Variability in discharge time also differs across solvers: Alt-Ergo and CVC5 show low standard deviations of 0.71ms, indicating stable performance across proof obligations, whereas CVC4 and Z3 exhibit higher variability with standard deviations of 2.12ms. These results indicate that, while proof success levels are similar across solvers, robustness and timing stability provide a clearer basis for differentiating prover behavior in practice.

\begin{table}[h]
\centering
\caption{The table reports the number of runs, observed proof success range, standard deviation of proof success, timeout counts, and Qed-time statistics of ACSL augmented C files by rule-based Python Script}
\label{tab:wp-summary}
\setlength{\tabcolsep}{4pt}

\resizebox{0.98\linewidth}{!}{%
\begin{tabular}{lccccccc}
\toprule
Solver & Runs & Min & Max & SD & TO & Qed & SD Qed \\
\midrule
Alt-Ergo & 115 & 75.0 & 100.0 & 4.56 & 6 & 7.50 & 0.71 \\
CVC4     & 113 & 75.0 & 100.0 & 4.64 & 0 & 8.50 & 2.12 \\
CVC5     & 111 & 75.0 & 100.0 & 4.72 & 0 & 7.50 & 0.71 \\
Z3       & 113 & 75.0 & 100.0 & 4.64 & 4 & 6.50 & 2.12 \\
\bottomrule
\end{tabular}
}
\end{table}

\begin{figure}[h!]
    \centering
    \begin{subfigure}{0.45\textwidth}
        \centering
        \includegraphics[width=\linewidth]{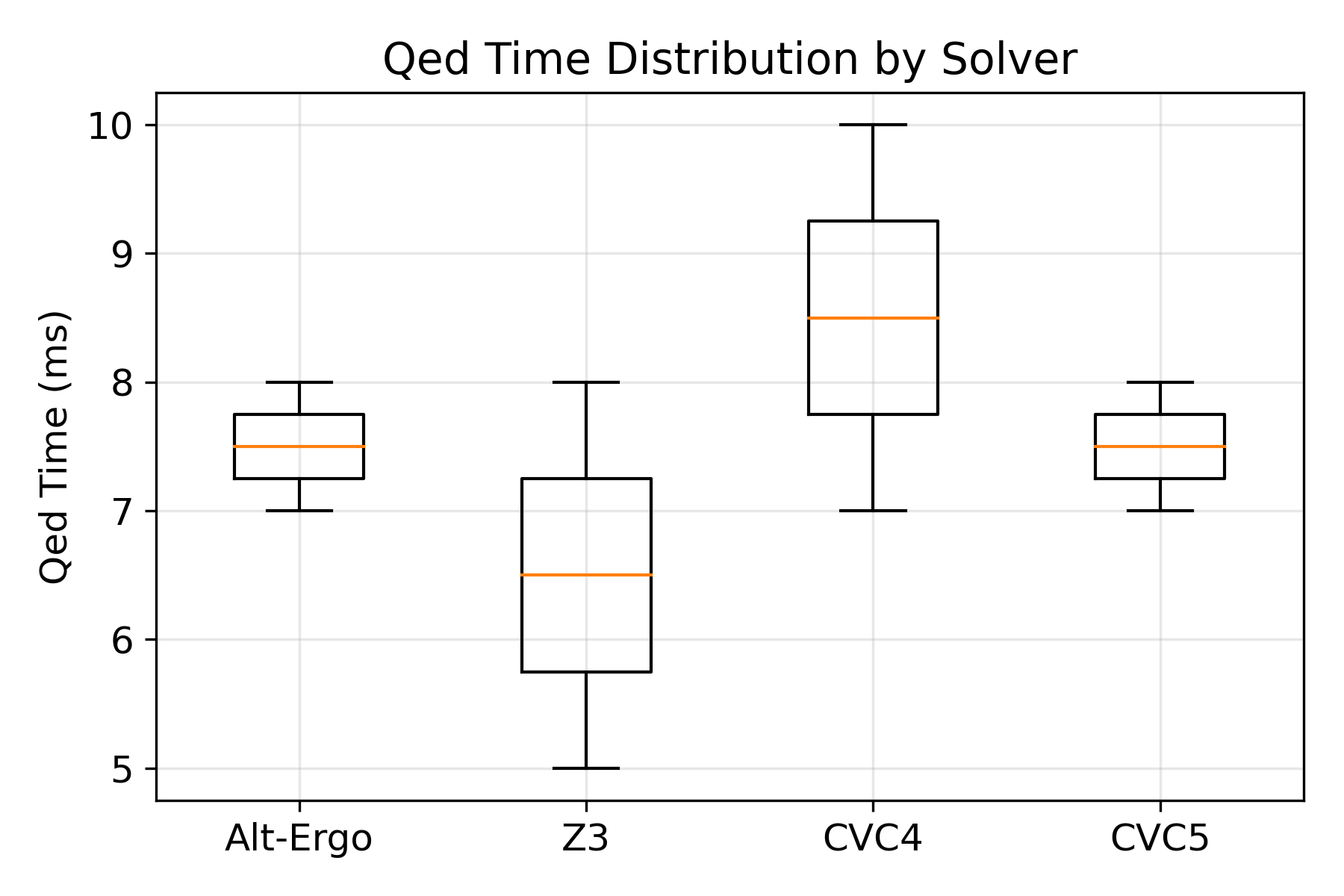}
        \caption{Qed Time Distribution}
        \label{fig:qed_time_boxplot_python}
    \end{subfigure}
    \hfill
    \begin{subfigure}{0.45\textwidth}
        \centering
        \includegraphics[width=\linewidth]{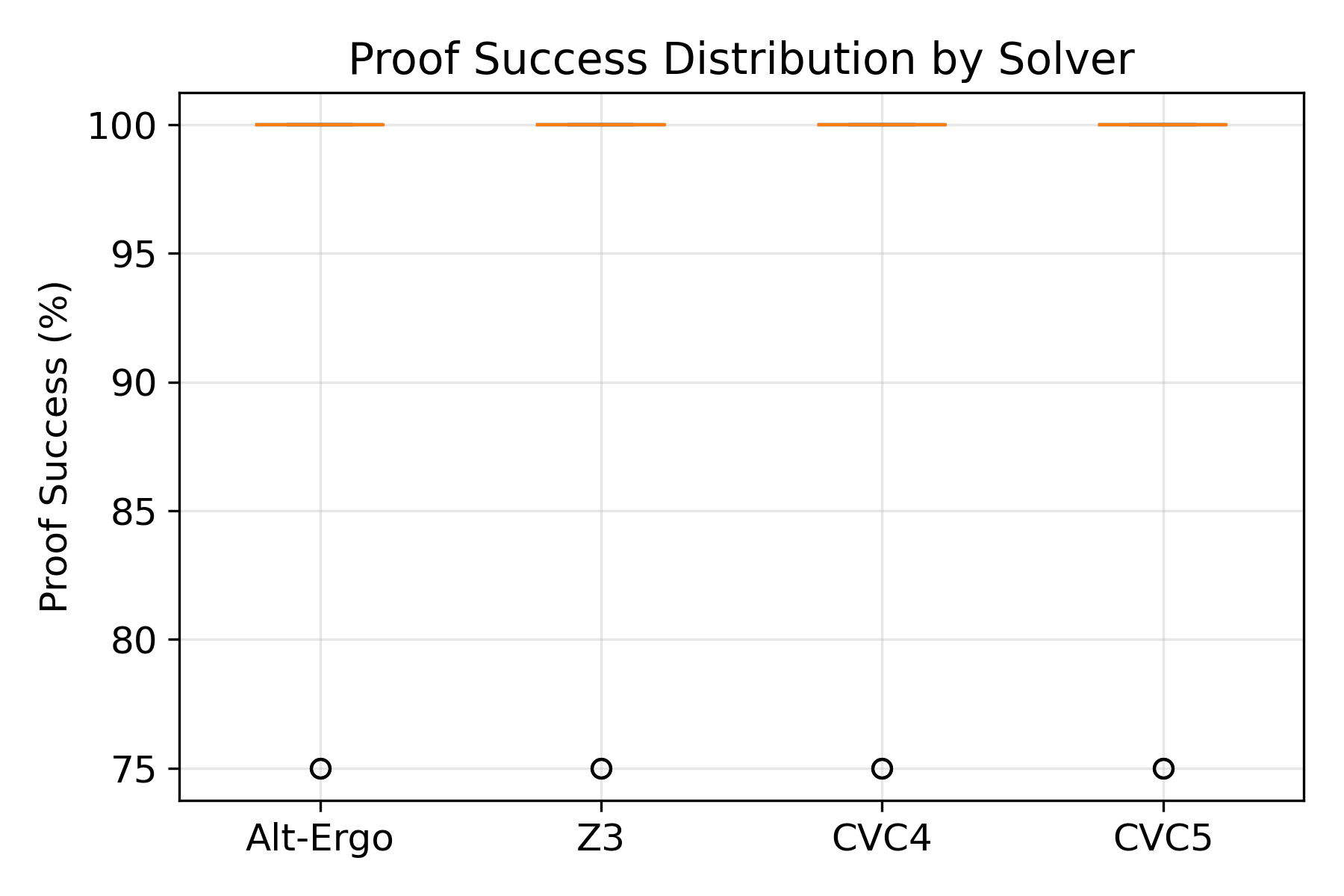}
        \caption{Proof Success Distribution}
        \label{fig:proof_success_boxplot_python}
    \end{subfigure}
    \caption{Solver Performance for Rule-based Python Script}
    \label{fig:solver_performance_python}
\end{figure}

Figure \ref{fig:qed_time_boxplot_python} presents a comparative analysis of Qed time variability, where the reported values correspond to the standard deviation of Qed times (in milliseconds) across four solvers, while Figure \ref{fig:proof_success_boxplot_python} presents the proof success distribution. Alt-Ergo and CVC5 exhibit similar variability profiles, with tightly clustered distributions and median standard deviation values around 7.5 ms, indicating relatively stable and consistent solver behavior. Z3 shows the lowest median standard deviation (approximately 6.5 ms), suggesting lower average variability; however, its wider interquartile range and extended whiskers indicate that variability fluctuates more significantly across instances, ranging from about 5 ms to 8 ms. In contrast, CVC4 demonstrates the highest median standard deviation (around 8.5 ms) and the largest overall spread, with values reaching up to roughly 10 ms, reflecting both higher variability and less predictable performance. Overall, the results highlight a trade-off between stability and dispersion across solvers: Z3 tends to exhibit lower central variability but with greater fluctuations, Alt-Ergo and CVC5 provide more consistent variability characteristics, and CVC4 shows the highest and most dispersed Qed time variability.

\subsection{Frama-C RTE Generation}

Table~\ref{tab:rte_wp_results} summarises 820 WP runs performed on ACSL specifications generated by Frama-C RTE, with each solver executed between 203 and 207 times. For all provers, proof success ranges from 30\% to 100\%, indicating that some verification tasks remain only partially discharged, while others are fully proved automatically. The dispersion of proof success is nearly identical across tools, with standard deviations tightly grouped around 25.8 (from 25.80 for Alt-Ergo to 25.89 for Z3), reflecting a comparable variability over the benchmark set. More substantial differences arise in timeout counts: Alt-Ergo records 366 timeouts, considerably more than Z3 (142), CVC5 (48), and CVC4 (42). In contrast, Qed-related metrics remain close, with mean values between 1.16 and 1.25 and standard deviations below 1.2. Overall, while proof coverage and variability appear consistent across solvers, timeout frequency constitutes the primary differentiating factor in these experiments.

\begin{table}[h!]
\centering
\resizebox{0.98\linewidth}{!}{
\begin{tabular}{lccccccc}
\toprule
Solver & Runs & Min & Max & SD & TO & Qed & SD Qed \\
\midrule
Alt-Ergo & 207 & 30.0 & 100.0 & 25.80 & 366 & 1.25 & 1.17 \\
CVC4     & 205 & 30.0 & 100.0 & 25.85 & 42  & 1.16 & 0.87 \\
CVC5     & 203 & 30.0 & 100.0 & 25.86 & 48  & 1.16 & 0.90 \\
Z3       & 205 & 30.0 & 100.0 & 25.89 & 142 & 1.22 & 0.89 \\
\bottomrule
\end{tabular}
}
\caption{WP verification results for ACSL specifications generated via Frama-C RTE}
\label{tab:rte_wp_results}
\end{table}

\begin{figure}[h!]
    \centering
    \begin{subfigure}{0.45\textwidth}
        \centering
        \includegraphics[width=\linewidth]{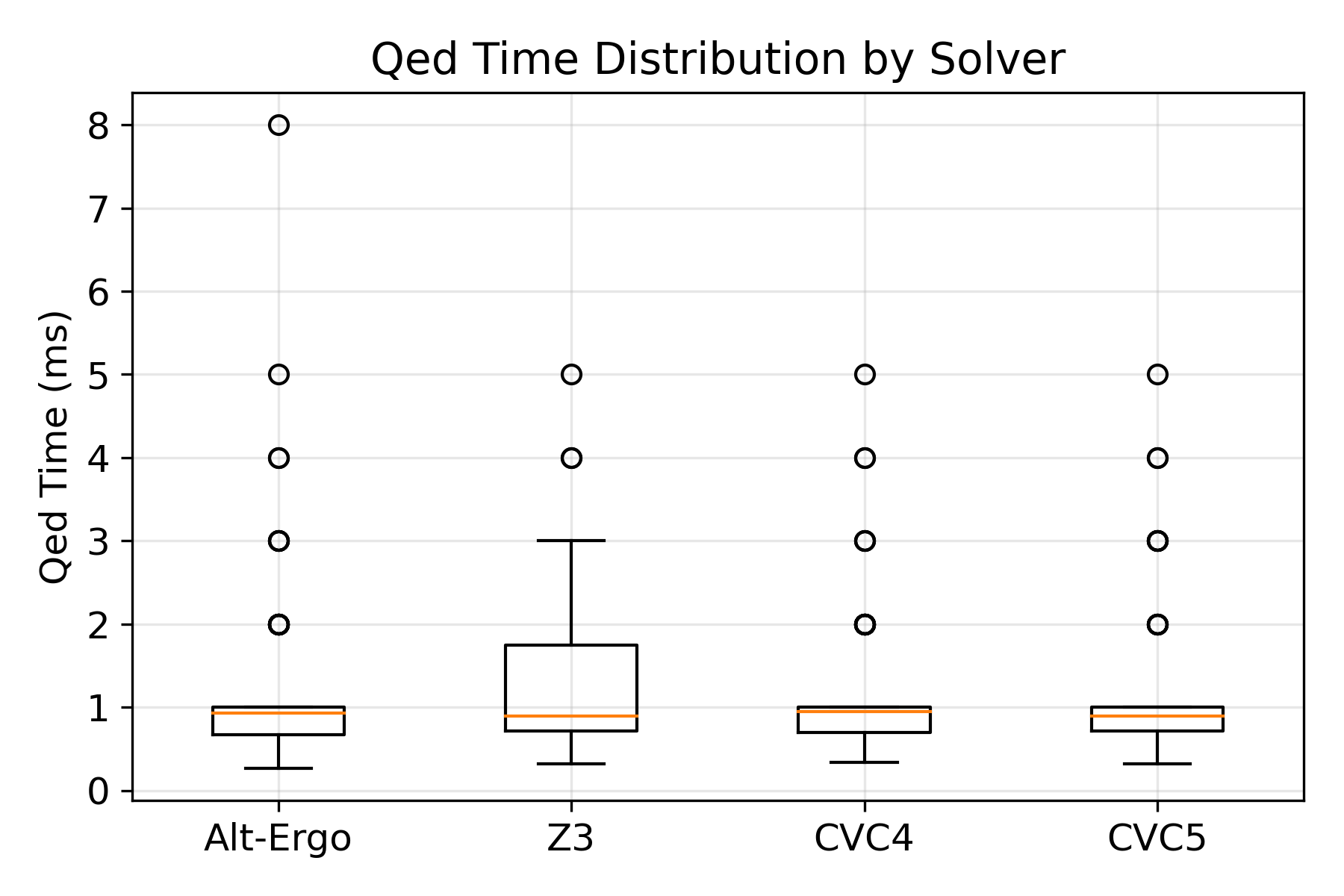}
        \caption{Qed Time Distribution}
        \label{fig:qed_time_boxplot_rte}
    \end{subfigure}
    \hfill
    \begin{subfigure}{0.45\textwidth}
        \centering
        \includegraphics[width=\linewidth]{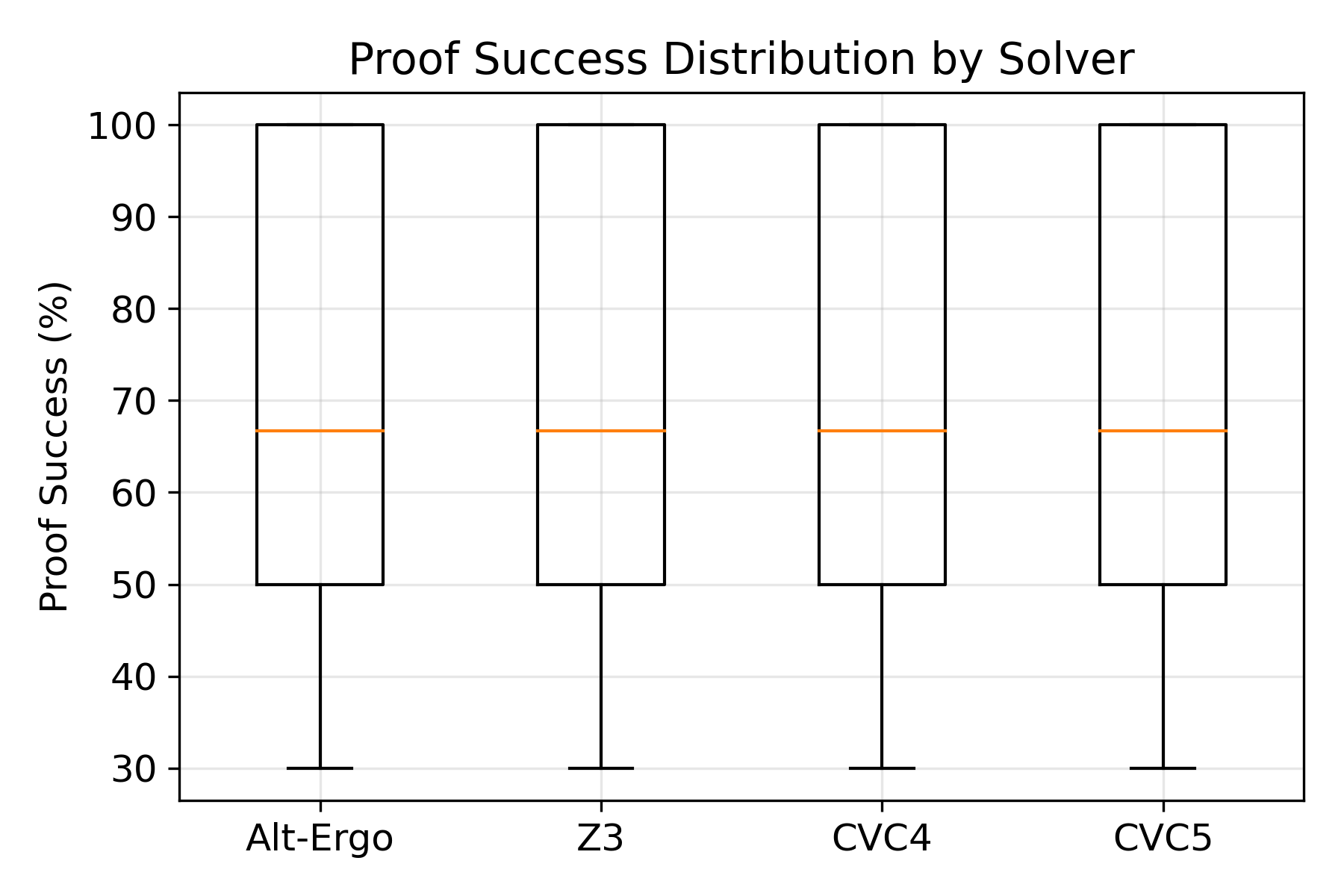}
        \caption{Proof Success Distribution}
        \label{fig:proof_success_boxplot_rte}
    \end{subfigure}
    \caption{Solver Performance for RTE-Generated ACSL}
    \label{fig:solver_performance_rte}
\end{figure}

Figure \ref{fig:qed_time_boxplot_rte} presents a comparative analysis of Qed time variability for ACSL specifications generated using the Frama-C RTE tool, where the reported values correspond to the standard deviation of Qed times (in milliseconds) across the four solvers, while Figure \ref{fig:proof_success_boxplot_rte} presents the proof success distribution. Overall, all solvers exhibit very low Qed time variability, reflecting the lightweight nature of proof obligations produced by RTE instrumentation. CVC4 and CVC5 show the most stable behavior, with median standard deviation values slightly below 1 ms and tightly clustered distributions, indicating highly consistent proof discharge times across verification runs. Z3 displays a comparable median variability (around 0.9 ms), though with a marginally wider interquartile range, suggesting slightly greater fluctuation across instances. Alt-Ergo exhibits the highest median standard deviation (approximately 1.17 ms) and a broader spread relative to the other solvers, indicating comparatively less stable Qed time behavior despite remaining within a low absolute range. Overall, the results highlight uniformly fast and stable Qed performance across solvers for RTE-generated ACSL, with CVC4 and CVC5 providing the most consistent profiles, Z3 showing moderate dispersion, and Alt-Ergo exhibiting the highest relative variability.

\subsection{GPT-5.2 Generation}

Table~\ref{tab:gpt5_wp_results} presents 700 WP executions over ACSL specifications generated by GPT-5.2, with each solver run between 173 and 177 times. Proof success spans from 16.67\% to 100\% for all provers, revealing a broader range of partial discharges compared to fully successful runs. The variability of results is again similar across tools, with standard deviations between 27.34 and 28.15, indicating a consistently wide dispersion over the benchmark set. Timeout behaviour, however, differs markedly: Alt-Ergo reports 272 timeouts, whereas Z3 records 20 and both CVC4 and CVC5 exhibit very few (8 and 10, respectively). Qed statistics show slightly higher averages than in the previous setting, ranging from 1.86 to 2.30, with relatively large standard deviations (up to 3.18), suggesting greater fluctuation in simplification effects. Overall, while proof coverage and dispersion remain comparable, timeout frequency and Qed variability distinguish solver performance in this configuration.

\begin{table}[h!]
\centering
\resizebox{0.98\linewidth}{!}{
\begin{tabular}{lccccccc}
\toprule
Solver & Runs & Min & Max & SD & TO & Qed & SD Qed \\
\midrule
Alt-Ergo & 177 & 16.67 & 100.0 & 28.15 & 272 & 1.86 & 2.88 \\
CVC4     & 175 & 16.67 & 100.0 & 27.64 & 8   & 2.00 & 2.88 \\
CVC5     & 173 & 16.67 & 100.0 & 27.34 & 10  & 2.30 & 3.18 \\
Z3       & 175 & 16.67 & 100.0 & 27.42 & 20  & 1.97 & 2.60 \\
\bottomrule
\end{tabular}
}
\caption{WP verification results for ACSL specifications generated by GPT-5.2}
\label{tab:gpt5_wp_results}
\end{table}

\begin{figure}[h!]
    \centering
    \begin{subfigure}{0.45\textwidth}
        \centering
        \includegraphics[width=\linewidth]{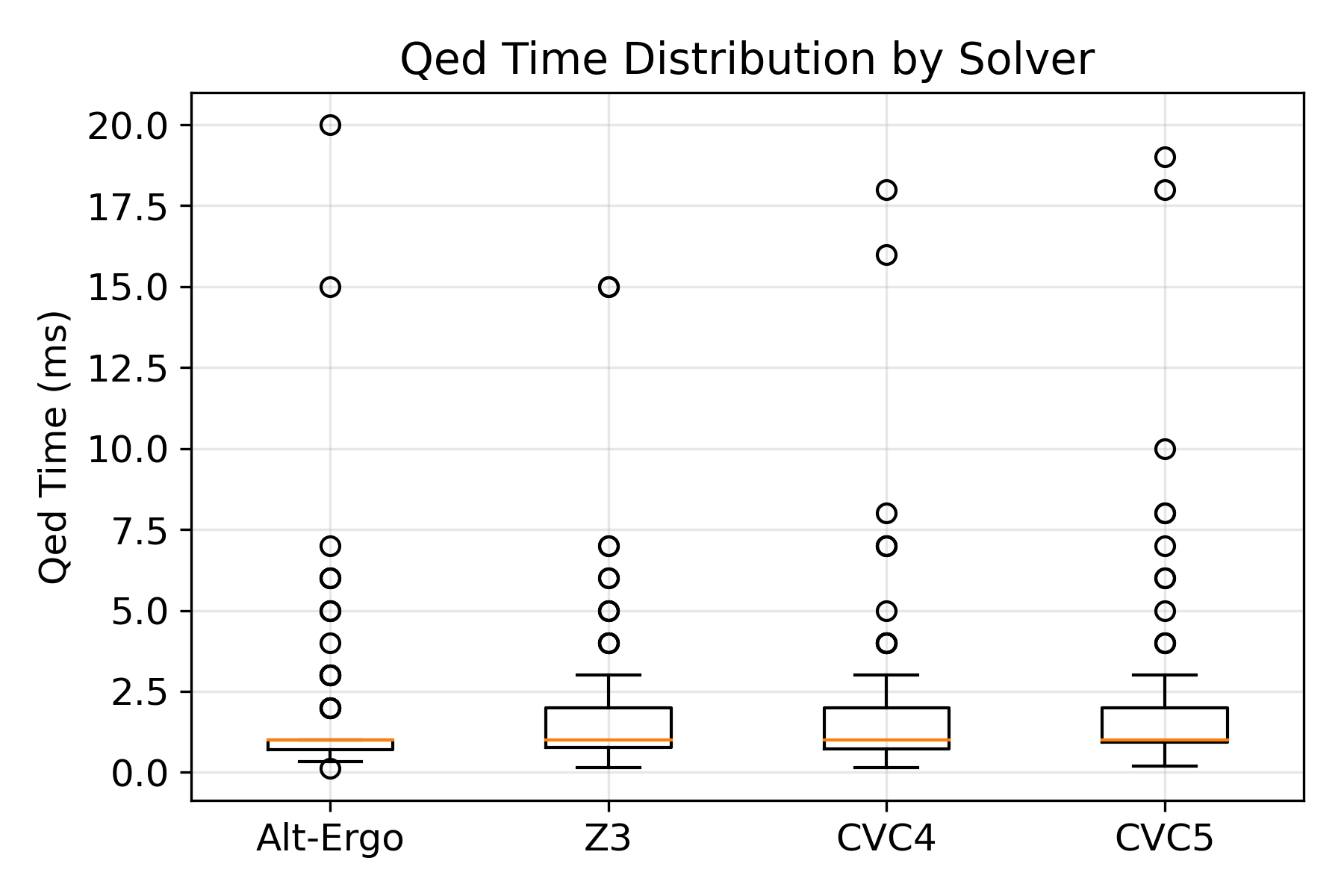}
        \caption{Qed Time Distribution}
        \label{fig:qed_time_boxplot_gpt5}
    \end{subfigure}
    \hfill
    \begin{subfigure}{0.45\textwidth}
        \centering
        \includegraphics[width=\linewidth]{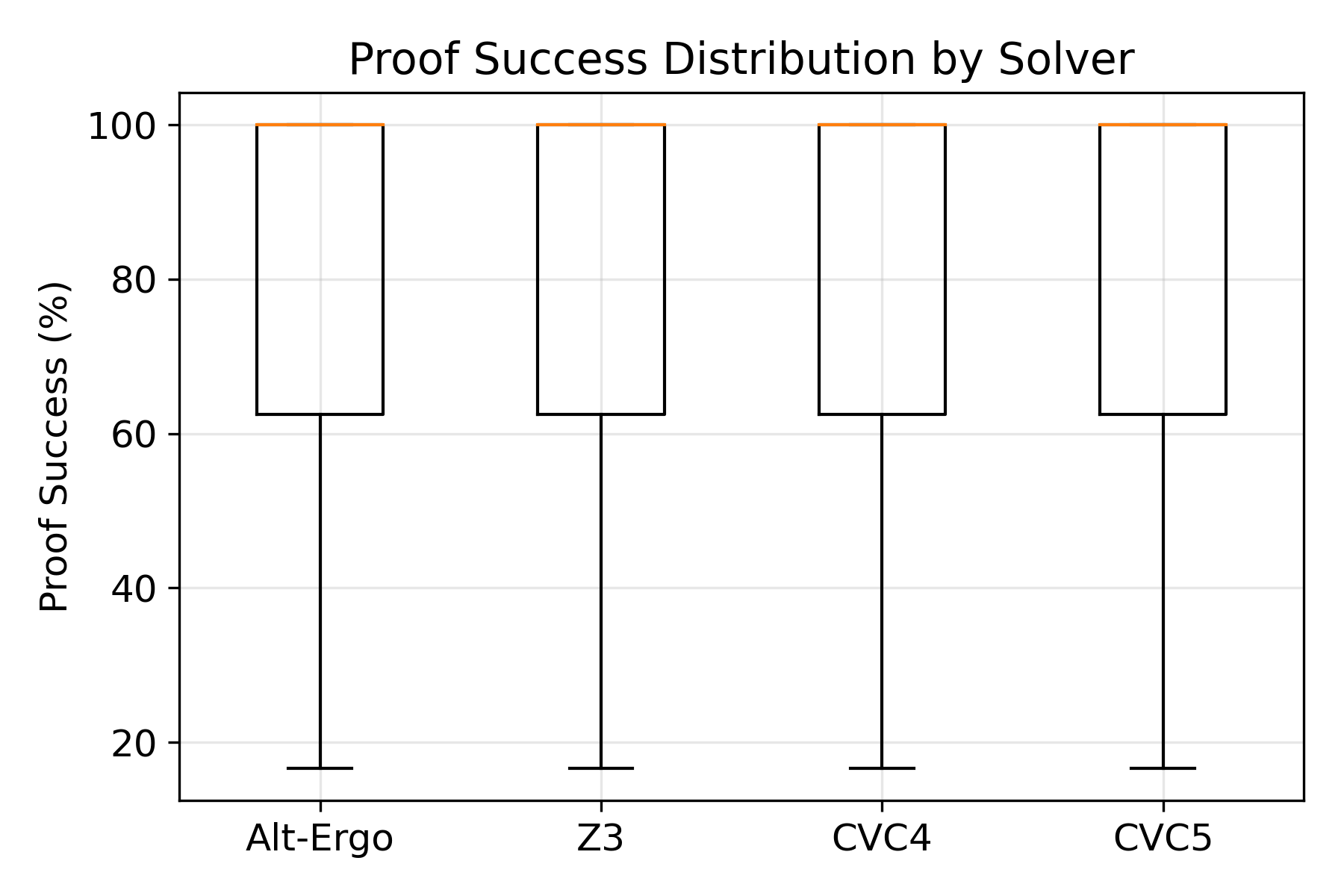}
        \caption{Proof Success Distribution}
        \label{fig:proof_success_boxplot_gpt5}
    \end{subfigure}
    \caption{Solver Performance for GPT-5--Generated ACSL}
    \label{fig:solver_performance_gpt5}
\end{figure}

Figure \ref{fig:qed_time_boxplot_gpt5} presents a comparative analysis of Qed time variability for ACSL specifications generated by the GPT-5 language model, where the reported values correspond to the standard deviation of Qed times (in milliseconds) across the four solvers, while Figure \ref{fig:proof_success_boxplot_gpt5} presents the proof success distribution. In contrast to RTE-generated specifications, all solvers exhibit substantially higher variability, reflecting the increased complexity and heterogeneity of LLM-generated ACSL annotations. Z3 shows the lowest median standard deviation (approximately 2.6 ms), indicating comparatively more stable Qed performance, although its distribution still spans a wide range of values, suggesting non-uniform behavior across verification instances. Alt-Ergo and CVC4 exhibit similar median variability levels (around 2.9 ms), with broader interquartile ranges that point to less predictable proof discharge times. CVC5 demonstrates the highest median standard deviation (approximately 3.2 ms) and the widest overall spread, indicating the most unstable Qed time behavior among the solvers. Overall, the results highlight a clear increase in Qed time dispersion for GPT-5–generated ACSL specifications: while Z3 offers the lowest central variability, Alt-Ergo and CVC4 show moderate instability, and CVC5 exhibits the highest and most dispersed variability profile.

\subsection{DeepSeek Generation}

Table~\ref{tab:deepseek_wp_results} aggregates 588 WP runs conducted on ACSL specifications generated by DeepSeek, with each solver executed between 145 and 149 times. For all provers, proof success ranges uniformly from 25\% to 100\%, indicating that while some benchmarks are only partially discharged, complete automation is frequently achieved. In contrast to the previous configurations, dispersion is notably lower, with standard deviations confined to a narrow interval between 13.97 and 14.72, suggesting more stable proof coverage across tasks. Timeout counts, however, remain uneven: Alt-Ergo records 130 timeouts, substantially exceeding Z3 (22), CVC5 (23), and especially CVC4 (13). Qed averages are considerably higher than in earlier settings, lying between 5.04 and 6.53, and are accompanied by large standard deviations (up to 20.67), reflecting significant variability in simplification impact. Overall, proof success appears more consistent, while timeout frequency and Qed dispersion differentiate solver behaviour.

\begin{table}[h!]
\centering
\resizebox{0.98\linewidth}{!}{
\begin{tabular}{lccccccc}
\toprule
Solver & Runs & Min & Max & SD & TO & Qed & SD Qed \\
\midrule
Alt-Ergo & 149 & 25.0 & 100.0 & 13.97 & 130 & 6.17 & 19.25 \\
CVC4     & 147 & 25.0 & 100.0 & 14.72 & 13  & 6.53 & 20.67 \\
CVC5     & 145 & 25.0 & 100.0 & 14.72 & 23  & 6.32 & 19.44 \\
Z3       & 147 & 25.0 & 100.0 & 14.30 & 22  & 5.04 & 16.27 \\
\bottomrule
\end{tabular}
}
\caption{WP verification results for ACSL specifications generated by DeepSeek}
\label{tab:deepseek_wp_results}
\end{table}

\begin{figure}[h!]
    \centering
    \begin{subfigure}{0.45\textwidth}
        \centering
        \includegraphics[width=\linewidth]{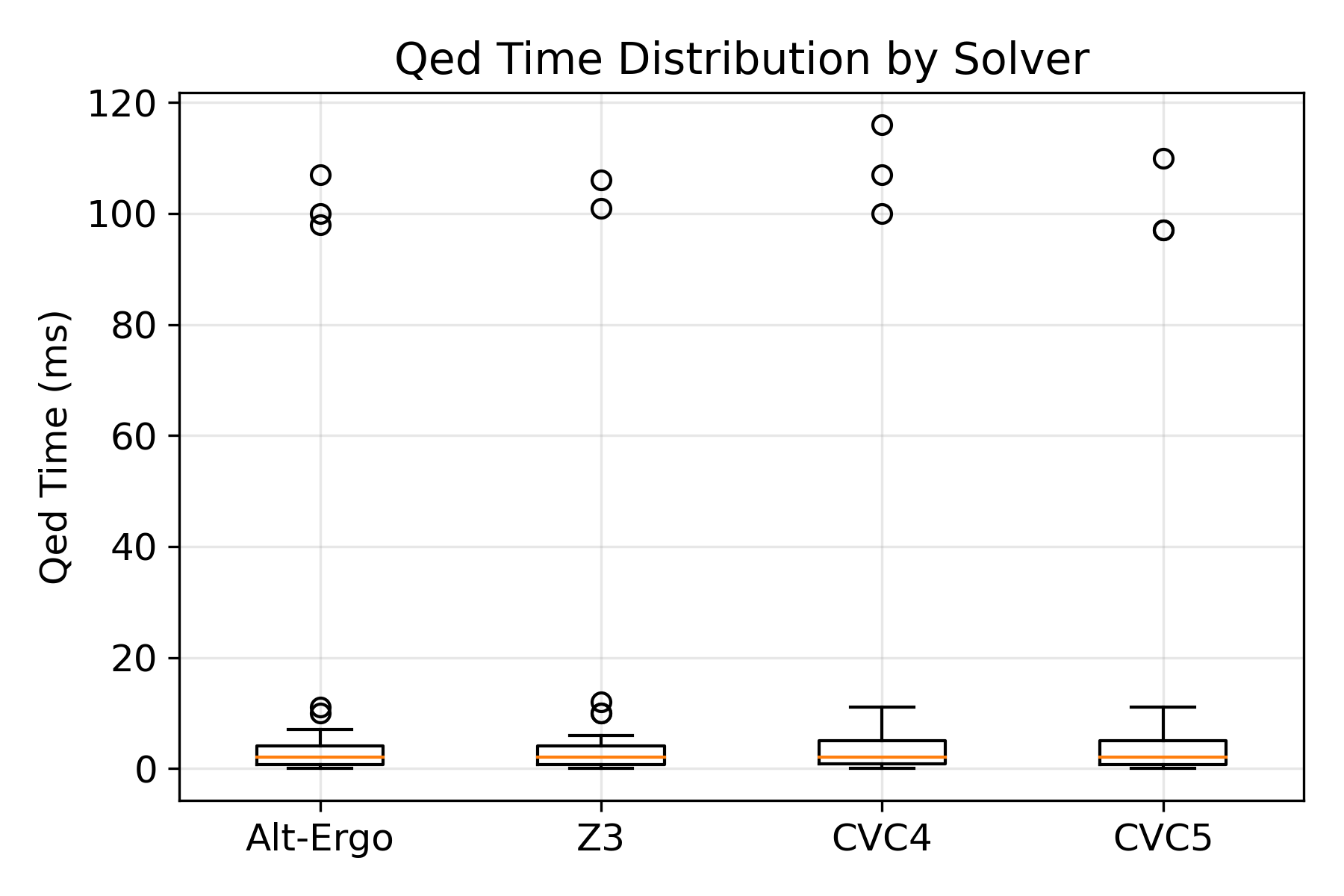}
        \caption{Qed Time Distribution}
        \label{fig:qed_time_boxplot_deepseek}
    \end{subfigure}
    \hfill
    \begin{subfigure}{0.45\textwidth}
        \centering
        \includegraphics[width=\linewidth]{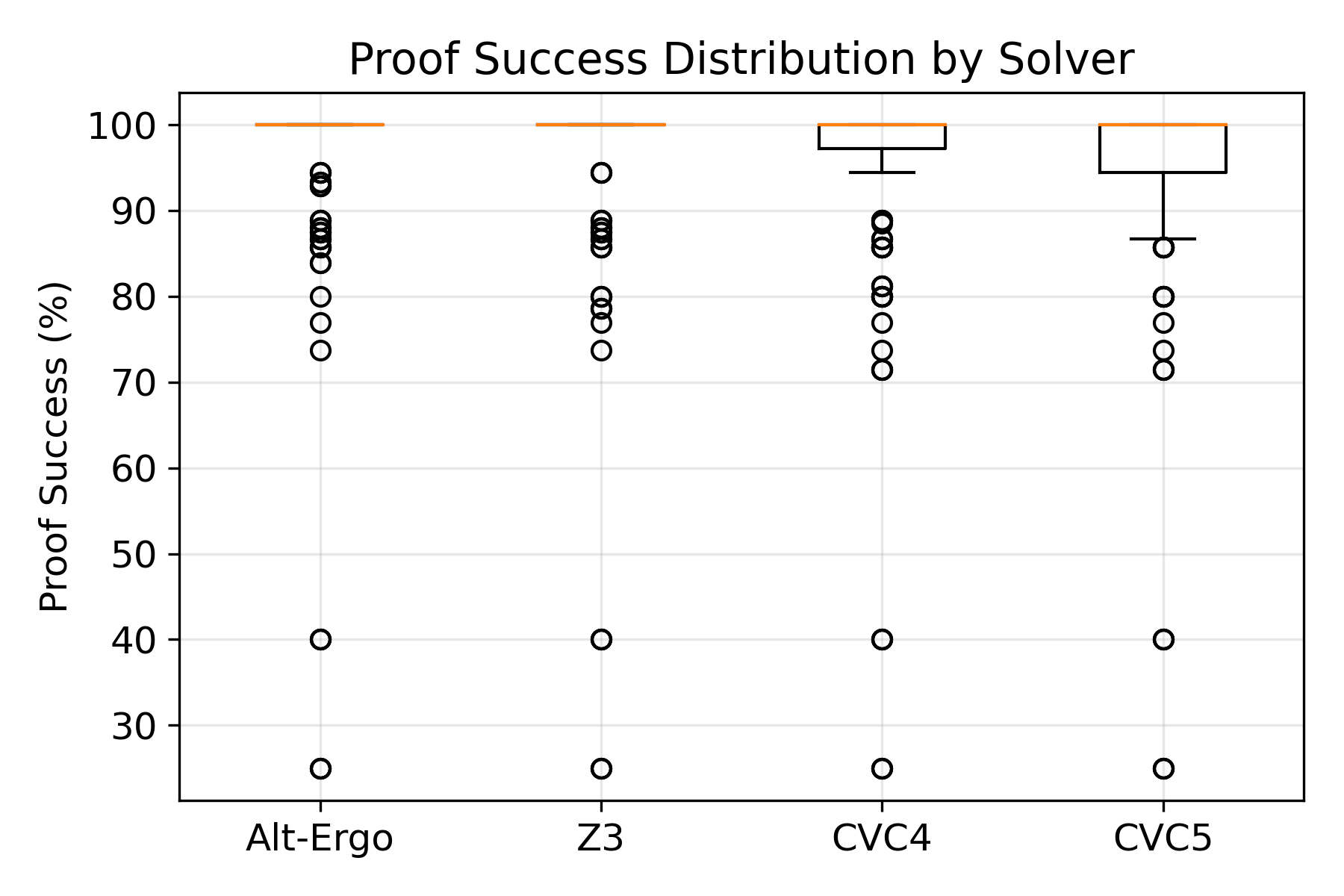}
        \caption{Proof Success Distribution}
        \label{fig:proof_success_boxplot_deepseek}
    \end{subfigure}
    \caption{Solver Performance for DeepSeek--Generated ACSL}
    \label{fig:solver_performance_deepseek}
\end{figure}

Figure \ref{fig:qed_time_boxplot_deepseek} presents a comparative analysis of Qed time variability for ACSL specifications generated by the DeepSeek model, where the reported values correspond to the standard deviation of Qed times (in milliseconds) across the four solvers, while Figure \ref{fig:proof_success_boxplot_deepseek} presents the proof success distribution. Compared to both RTE-generated and GPT-5–generated ACSL, DeepSeek specifications result in substantially higher Qed time dispersion, indicating significantly less predictable solver behavior. Z3 exhibits the lowest median standard deviation (approximately 16.3 ms), suggesting comparatively more stable proof discharge times, in this more challenging verification condition. Alt-Ergo, CVC4, and CVC5 show markedly higher variability, with median standard deviation values clustered around 19–21 ms and wide interquartile ranges. Among these, CVC4 demonstrates the highest median variability and the broadest spread, indicating the most unstable Qed time behavior, while Alt-Ergo and CVC5 display similarly high dispersion with slightly lower central values. Overall, the results highlight a pronounced increase in Qed time variability for DeepSeek-generated ACSL specifications: Z3 offers the lowest central variability, whereas Alt-Ergo and CVC5 show high but comparable instability, and CVC4 exhibits the highest and most dispersed variability profile among the solvers.

\subsection{OLMo3 Generation}

Table~\ref{tab:olmo3_wp_results} summarises 532 WP runs on ACSL specifications generated by OLMo3, with each solver invoked between 132 and 134 times. All provers share identical proof-success bounds, ranging from 33.33\% to 100\%, and exhibit exactly the same standard deviation (25.56), indicating an indistinguishable dispersion of results across the benchmark suite. Thus, in terms of coverage and variability, the four solvers behave almost identically in this setting. Differences arise primarily in timeout counts: CVC4 and CVC5 complete all runs without timeouts, whereas Alt-Ergo and Z3 report 92 and 77 timeouts, respectively. Qed averages remain close, between 2.81 and 3.05, with modest standard deviations (1.24–1.56), suggesting limited variation in simplification effects. Overall, while proof success statistics are perfectly aligned across solvers, timeout behaviour constitutes the principal factor separating their performance on these specifications.

\begin{table}[h!]
\centering
\resizebox{0.98\linewidth}{!}{
\begin{tabular}{lccccccc}
\toprule
Solver & Runs & Min & Max & SD & TO & Qed & SD Qed \\
\midrule
Alt-Ergo & 134 & 33.33 & 100.0 & 25.56 & 92 & 2.92 & 1.24 \\
CVC4     & 133 & 33.33 & 100.0 & 25.56 & 0  & 2.81 & 1.25 \\
CVC5     & 132 & 33.33 & 100.0 & 25.56 & 0  & 3.05 & 1.47 \\
Z3       & 133 & 33.33 & 100.0 & 25.56 & 77 & 2.87 & 1.56 \\
\bottomrule
\end{tabular}
}
\caption{WP verification results for ACSL specifications generated by OLMo3}
\label{tab:olmo3_wp_results}
\end{table}

\begin{figure}[h!]
    \centering
    \begin{subfigure}{0.45\textwidth}
        \centering
        \includegraphics[width=\linewidth]{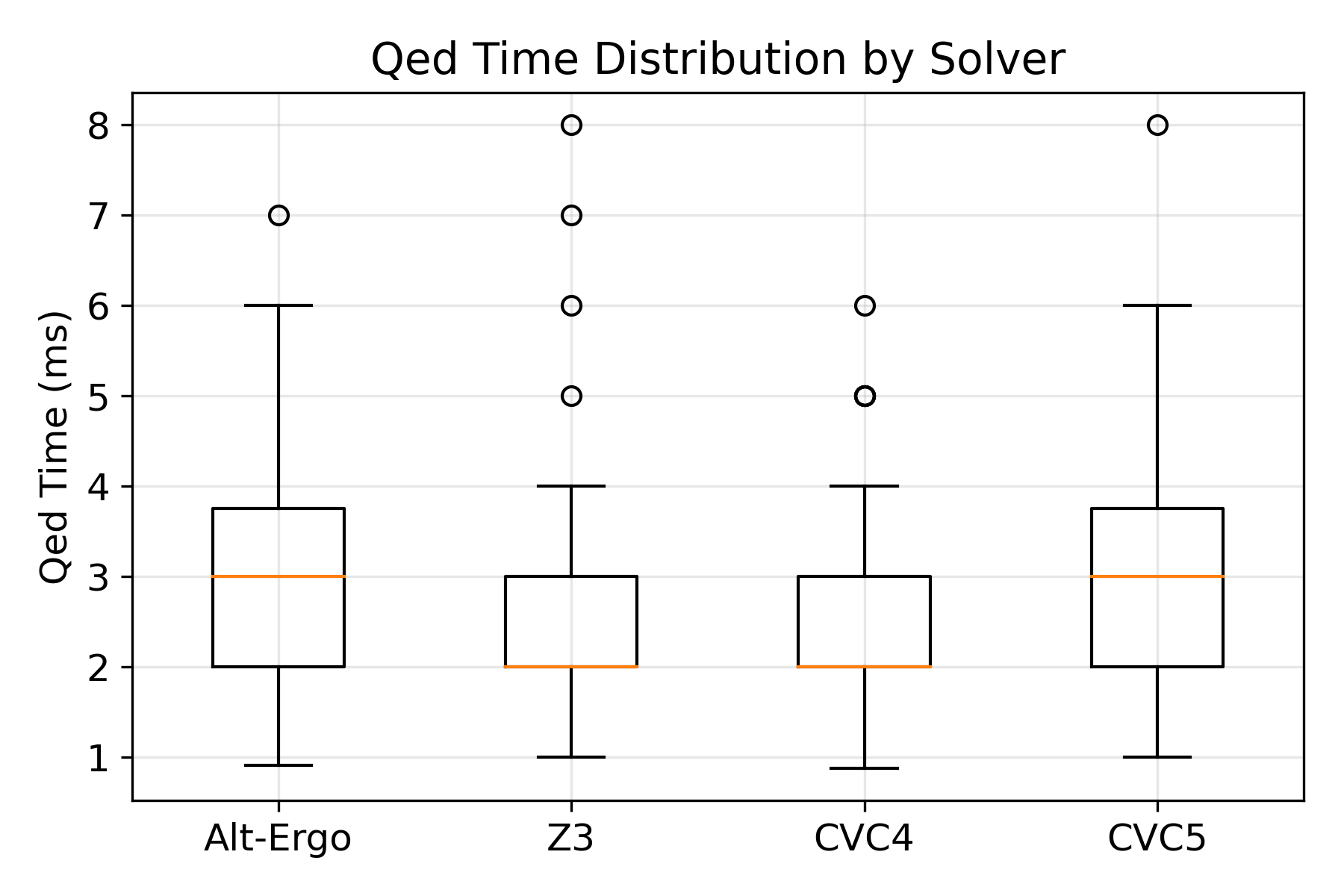}
        \caption{Qed Time Distribution}
        \label{fig:qed_time_boxplot_olmo3}
    \end{subfigure}
    \hfill
    \begin{subfigure}{0.45\textwidth}
        \centering
        \includegraphics[width=\linewidth]{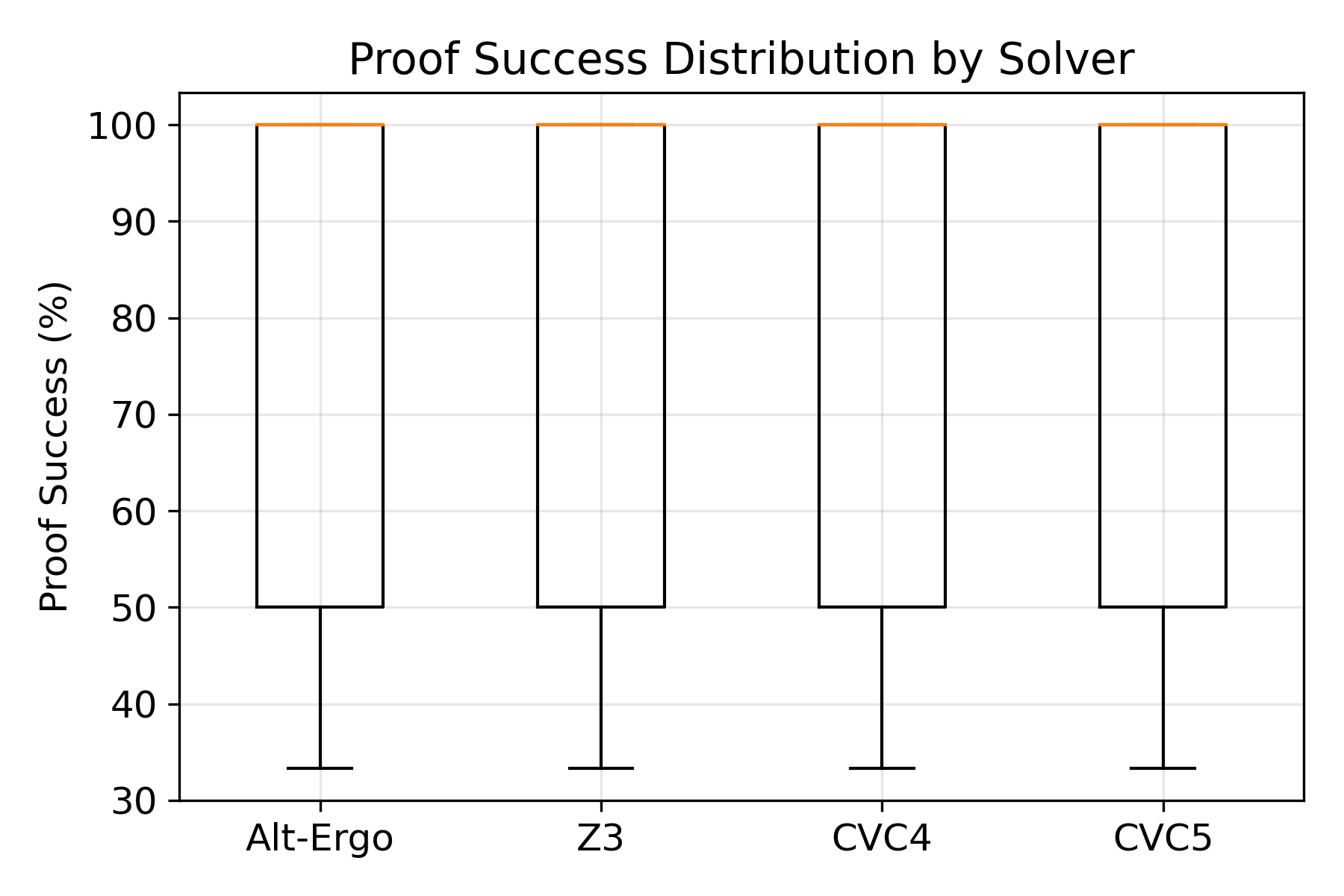}
        \caption{Proof Success Distribution}
        \label{fig:proof_success_boxplot_olmo3}
    \end{subfigure}
    \caption{Solver Performance for OLMo3--Generated ACSL}
    \label{fig:solver_performance_olmo3}
\end{figure}

Figure \ref{fig:qed_time_boxplot_olmo3} presents a comparative analysis of Qed time variability for ACSL specifications generated by the OLMo3 model, where the reported values correspond to the standard deviation of Qed times (in milliseconds) across the four solvers, while the Figure \ref{fig:proof_success_boxplot_olmo3} presents the proof success distribution. In contrast to DeepSeek- and GPT-5–generated specifications, OLMo3 exhibits markedly lower Qed time dispersion, indicating more stable solver behavior. Alt-Ergo and CVC4 show very similar median standard deviation values (around 1.2–1.3 ms) with tightly clustered distributions, reflecting consistent proof discharge performance across verification runs. CVC5 displays a slightly higher median variability (approximately 1.5 ms) and a moderately wider interquartile range, suggesting limited additional fluctuation. Z3 exhibits the highest median standard deviation among the solvers (around 1.6 ms), accompanied by a broader spread, indicating comparatively less stable behavior, though still within a relatively low absolute range. Overall, the results highlight that ACSL specifications generated by OLMo3 lead to stable and predictable Qed time behavior across solvers: Alt-Ergo and CVC4 provide the most consistent variability profiles, CVC5 shows moderate dispersion, and Z3 exhibits the highest, yet still limited, variability.

Across the five ACSL generation techniques, a clear gradient in verification stability emerges. Rule-based and RTE-generated ACSL consistently produce the lowest Qed time variability across all solvers, reflecting the uniform and predictable structure of the generated proof obligations. OLMo3 occupies a middle ground, maintaining low variability while introducing moderately richer specifications that remain solver friendly. In contrast, GPT-5 and DeepSeek generate substantially more heterogeneous ACSL, leading to sharply increased Qed time dispersion and less predictable verification behavior, particularly evident in the wide spreads observed across solvers.

From a solver perspective, Z3 generally maintains consistent runtime performance on complex, LLM-generated ACSL but exhibits broader dispersion, indicating sensitivity to specification irregularities. Alt-Ergo and CVC5 remain stable on structured techniques but degrade under more expressive LLM outputs, while CVC4 shows the strongest consistency on simpler inputs and the most pronounced variability on complex ones. Overall, the comparison highlights a fundamental trade-off: increased expressiveness in ACSL generation correlates with reduced solver stability, whereas disciplined, structure-driven techniques yield more reliable and predictable verification outcomes.

Overall, our results reveal a clear trade-off between expressiveness and verification stability. Tool-generated ACSL (rule-based and RTE) delivers fast, predictable, and highly consistent solver performance, with low mean and median Qed times and minimal dispersion. OLMo3-generated specifications maintain this stability while introducing modest complexity. In contrast, GPT-5 and DeepSeek-generated ACSL produce more heterogeneous proof obligations, leading to higher and more variable Qed times, wider distributions, and increased solver sensitivity. Among solvers, CVC4 and CVC5 provide the most consistent behavior, Z3 achieves lower central variability but exhibits broader fluctuations on complex inputs, and Alt-Ergo experiences the highest timeout rates. These findings emphasize that achieving reliable large-scale verification requires balancing specification richness with controlled variability to ensure predictable solver performance across diverse ACSL generation techniques.

\section{Conclusions}
\label{sec:conclusion}

This paper presented an empirical evaluation of five ACSL annotation generation strategies, including rule-based, tool-supported, and LLM-based approaches, within a unified verification framework. Using the Frama-C WP plugin and multiple SMT solvers, we assessed their effectiveness in terms of proof success, solver behavior, and processing characteristics. Our results show that rule-based approaches and the Frama-C RTE plugin remain more reliable for producing verifiable annotations, particularly for safety-related properties. In contrast, LLM-based methods demonstrate more variable performance, generating a broader range of annotations but with less consistency in verification outcomes. These findings suggest that, while LLMs are not yet a replacement for established rule-based techniques, they offer potential as complementary tools for expanding specification coverage. Future work includes improving the reliability of LLM-generated annotations and exploring hybrid approaches that combine symbolic and learning-based methods.

\subsection{Lessons Learned}
\label{subsec:lessonslearned}

Our empirical evaluation highlights several practical insights for the use of automated ACSL annotation generation strategies.

\paragraph{Reliability of rule-based approaches.}
Rule-based methods and the Frama-C RTE plugin provide the most consistent verification performance across benchmarks. Their focus on well-defined safety properties makes them a reliable baseline, particularly in safety-critical contexts where predictable behavior is essential.

\paragraph{Variability of LLM-based approaches.}
LLM-generated annotations exhibit greater variability in both structure and verification success. While they can produce a broader range of annotations, including functional specifications, their outputs are less consistent and may include redundant or weak constraints.

\paragraph{When LLMs are useful.}
LLM-based approaches are most promising in scenarios where broader specification coverage is desired, especially when manual specification effort is limited. However, their outputs currently require careful validation before integration into formal verification workflows.

\paragraph{Differences across LLMs.}
Among the evaluated models, DeepSeek-V3.2 demonstrates more stable performance compared to GPT-5.2 and OLMo 3.1. A possible explanation is its stronger capability in structured code reasoning, although a detailed investigation of this behavior is left for future work.

\paragraph{Practical implication.}
Overall, our results suggest that LLM-based annotation generation should be viewed as complementary to existing rule-based techniques rather than a replacement. Combining the reliability of rule-based methods with the flexibility of LLMs represents a promising direction for future research.

\section*{Use of AI-Assisted Tools}

We acknowledge the use of free version of GPT-5.2 for refining the textual presentation of the paper.
The model was applied to improve clarity and coherence. Followed by this, the text has been thoroughly
reviewed and discussed by all authors to ensure accuracy and integrity.

\begin{acks}
This work has been supported by the European Commission as part of the ADAPT Centre for Digital Content Technology which is funded under the SFI Research Centres Programme (Grant 13/RC/2106) and is co-funded under the European Regional Development Fund.
\end{acks}

\bibliographystyle{ACM-Reference-Format}
\balance
\bibliography{ftfjp2026Bib}

@Misc{cosler2023nl2specinteractivelytranslatingunstructured,
  author        = {Matthias Cosler and Christopher Hahn and Daniel Mendoza and Frederik Schmitt and Caroline Trippel},
  title         = {nl2spec: Interactively Translating Unstructured Natural Language to Temporal Logics with Large Language Models},
  year          = {2023},
  archiveprefix = {arXiv},
  eprint        = {2303.04864},
  primaryclass  = {cs.LO},
  url           = {https://arxiv.org/abs/2303.04864},
}

@InProceedings{10691792,
  author    = {Fang, Wenji and Li, Mengming and Li, Min and Yan, Zhiyuan and Liu, Shang and Zhang, Hongce and Xie, Zhiyao},
  booktitle = {2024 IEEE LLM Aided Design Workshop (LAD)},
  title     = {AssertLLM: Generating Hardware Verification Assertions from Design Specifications via Multi-LLMs},
  year      = {2024},
  pages     = {1-1},
  doi       = {10.1109/LAD62341.2024.10691792},
}

@InProceedings{Req2SpecPaper,
  author    = {Nayak, Anmol and Timmapathini, Hari Prasad and Murali, Vidhya and Ponnalagu, Karthikeyan and Venkoparao, Vijendran Gopalan and Post, Amalinda},
  booktitle = {Requirements Engineering: Foundation for Software Quality: 28th International Working Conference, REFSQ 2022, Birmingham, UK, March 21–24, 2022, Proceedings},
  title     = {Req2Spec: Transforming Software Requirements into Formal Specifications Using Natural Language Processing},
  year      = {2022},
  address   = {Berlin, Heidelberg},
  pages     = {87–95},
  publisher = {Springer-Verlag},
  doi       = {},
  isbn      = {978-3-030-98463-2},
  location  = {Birmingham, United Kingdom},
  numpages  = {9},
  url       = {},
}

@Misc{fan2025evaluatingabilitylargelanguage,
  author        = {Wen Fan and Marilyn Rego and Xin Hu and Sanya Dod and Zhaorui Ni and Danning Xie and Jenna DiVincenzo and Lin Tan},
  title         = {Evaluating the Ability of Large Language Models to Generate Verifiable Specifications in VeriFast},
  year          = {2025},
  archiveprefix = {arXiv},
  eprint        = {2411.02318},
  primaryclass  = {cs.SE},
  url           = {https://arxiv.org/abs/2411.02318},
}

@Misc{LemurWu2024,
  author        = {Haoze Wu and Clark Barrett and Nina Narodytska},
  title         = {Lemur: Integrating Large Language Models in Automated Program Verification},
  year          = {2024},
  archiveprefix = {arXiv},
  eprint        = {2310.04870},
  primaryclass  = {cs.FL},
  url           = {https://arxiv.org/abs/2310.04870},
}

@InProceedings{Granberry2025,
  author    = {Granberry, George and Ahrendt, Wolfgang and Johansson, Moa},
  booktitle = {Leveraging Applications of Formal Methods, Verification and Validation. Specification and Verification},
  title     = {Towards Integrating Copiloting and Formal Methods},
  year      = {2025},
  address   = {Cham},
  editor    = {Margaria, Tiziana and Steffen, Bernhard},
  pages     = {144--158},
  publisher = {Springer Nature Switzerland},
  isbn      = {978-3-031-75380-0},
}

@InProceedings{Granberry2025a,
  author    = {Granberry, George and Ahrendt, Wolfgang and Johansson, Moa},
  booktitle = {Integrated Formal Methods},
  title     = {Specify What? Enhancing Neural Specification Synthesis by Symbolic Methods},
  year      = {2025},
  address   = {Cham},
  editor    = {Kosmatov, Nikolai and Kov{\'a}cs, Laura},
  pages     = {307--325},
  publisher = {Springer Nature Switzerland},
  isbn      = {978-3-031-76554-4},
}

@Article{Ma2024,
  author        = {Lezhi Ma and Shangqing Liu and Yi Li and Xiaofei Xie and Lei Bu},
  title         = {SpecGen: Automated Generation of Formal Program Specifications via Large Language Models},
  year          = {2024},
  archiveprefix = {arXiv},
  eprint        = {2401.08807},
  primaryclass  = {cs.SE},
  url           = {https://arxiv.org/abs/2401.08807},
}

@InProceedings{MarieFarrell202507,
  author    = {Weiqi Wang and Marie Farrell and Lucas C. Cordeiro and Liping Zhao},
  booktitle = {33rd {IEEE} International Requirements Engineering Conference, {RE} 2025, Valencia, Spain, September 1-5, 2025},
  title     = {Supporting Software Formal Verification with Large Language Models: An Experimental Study},
  year      = {2025},
  pages     = {423--431},
  publisher = {{IEEE}},
  bibsource = {dblp computer science bibliography, https://dblp.org},
  doi       = {10.1109/RE63999.2025.00049},
  url       = {https://doi.org/10.1109/RE63999.2025.00049},
}

@InProceedings{PALMforCoqProofs2024,
  author    = {Minghai Lu and Benjamin Delaware and Tianyi Zhang},
  booktitle = {Proceedings of the 39th {IEEE/ACM} International Conference on Automated Software Engineering, {ASE} 2024, Sacramento, CA, USA, October 27 - November 1, 2024},
  title     = {Proof Automation with Large Language Models},
  year      = {2024},
  editor    = {Vladimir Filkov and Baishakhi Ray and Minghui Zhou},
  pages     = {1509--1520},
  publisher = {{ACM}},
  bibsource = {dblp computer science bibliography, https://dblp.org},
  doi       = {10.1145/3691620.3695521},
  url       = {https://doi.org/10.1145/3691620.3695521},
}

@InProceedings{LLMs4OREZhao202503,
  author    = {Yihang Zhao},
  booktitle = {The Semantic Web: {ESWC} 2025 Satellite Events - Portoroz, Slovenia, June 1-5, 2025, Proceedings},
  title     = {Leveraging Large Language Models for Ontology Requirements Engineering},
  year      = {2025},
  editor    = {Edward Curry and Valentina Presutti and John P. McCrae and Mehwish Alam and Pieter Colpaert and Josiane Xavier Parreira and Diego Collarana and Marta Sabou and Andreas Harth and Pasquale Lisena},
  pages     = {254--264},
  publisher = {Springer},
  series    = {Lecture Notes in Computer Science},
  volume    = {15832},
  bibsource = {dblp computer science bibliography, https://dblp.org},
  doi       = {10.1007/978-3-031-99554-5\_40},
  url       = {https://doi.org/10.1007/978-3-031-99554-5\_40},
}

@InProceedings{AutoSpecCAV2024,
  author    = {Cheng Wen and Jialun Cao and Jie Su and Zhiwu Xu and Shengchao Qin and Mengda He and Haokun Li and Shing{-}Chi Cheung and Cong Tian},
  booktitle = {Computer Aided Verification - 36th International Conference, {CAV} 2024, Montreal, QC, Canada, July 24-27, 2024, Proceedings, Part {II}},
  title     = {Enchanting Program Specification Synthesis by Large Language Models Using Static Analysis and Program Verification},
  year      = {2024},
  editor    = {Arie Gurfinkel and Vijay Ganesh},
  pages     = {302--328},
  publisher = {Springer},
  series    = {Lecture Notes in Computer Science},
  volume    = {14682},
  bibsource = {dblp computer science bibliography, https://dblp.org},
  doi       = {10.1007/978-3-031-65630-9\_16},
  url       = {https://doi.org/10.1007/978-3-031-65630-9\_16},
}

@Article{FindingLoopInvariantsArxiv2023,
  author     = {Adharsh Kamath and Aditya Senthilnathan and Saikat Chakraborty and Pantazis Deligiannis and Shuvendu K. Lahiri and Akash Lal and Aseem Rastogi and Subhajit Roy and Rahul Sharma},
  journal    = {CoRR},
  title      = {Finding Inductive Loop Invariants using Large Language Models},
  year       = {2023},
  volume     = {abs/2311.07948},
  bibsource  = {dblp computer science bibliography, https://dblp.org},
  doi        = {10.48550/ARXIV.2311.07948},
  eprint     = {2311.07948},
  eprinttype = {arXiv},
  url        = {https://doi.org/10.48550/arXiv.2311.07948},
}

@Article{HilbertArXiV202509,
  author     = {Sumanth Varambally and Thomas Voice and Yanchao Sun and Zhifeng Chen and Rose Yu and Ke Ye},
  journal    = {CoRR},
  title      = {Hilbert: Recursively Building Formal Proofs with Informal Reasoning},
  year       = {2025},
  volume     = {abs/2509.22819},
  bibsource  = {dblp computer science bibliography, https://dblp.org},
  doi        = {10.48550/ARXIV.2509.22819},
  eprint     = {2509.22819},
  eprinttype = {arXiv},
  url        = {https://doi.org/10.48550/arXiv.2509.22819},
}

@Article{RvLLM202505,
  author     = {Yedi Zhang and Sun Yi Emma and Annabelle Lee Jia En and Jin Song Dong},
  journal    = {CoRR},
  title      = {RvLLM: {LLM} Runtime Verification with Domain Knowledge},
  year       = {2025},
  volume     = {abs/2505.18585},
  bibsource  = {dblp computer science bibliography, https://dblp.org},
  doi        = {10.48550/ARXIV.2505.18585},
  eprint     = {2505.18585},
  eprinttype = {arXiv},
  url        = {https://doi.org/10.48550/arXiv.2505.18585},
}

@Article{SVLLM4SoCDesign202506,
  author        = {Dipayan Saha and Shams Tarek and Hasan Al Shaikh and Khan Thamid Hasan and Pavan Sai Nalluri and Md. Ajoad Hasan and Nashmin Alam and Jingbo Zhou and Sujan Kumar Saha and Mark Tehranipoor and Farimah Farahmandi},
  title         = {SV-LLM: An Agentic Approach for SoC Security Verification using Large Language Models},
  year          = {2025},
  archiveprefix = {arXiv},
  eprint        = {2506.20415},
  primaryclass  = {cs.CR},
  url           = {https://arxiv.org/abs/2506.20415},
}

@InProceedings{BegEtAl2025,
  author    = {Beg, Arshad and O'Donoghue, Diarmuid and Monahan, Rosemary},
  booktitle = {Proceedings of OVERLAY 2025: Artificial Intelligence and Formal Verification, Logic, Automata, and Synthesis},
  title     = {Leveraging {LLM}s for Formal Software Requirements: Challenges and Prospects},
  year      = {2025},
  pages     = {95--105},
  publisher = {CEUR-WS.org},
  series    = {CEUR Workshop Proceedings},
  volume    = {4142},
  issn      = {1613-0073},
  url       = {https://ceur-ws.org/Vol-4142/paper11.pdf},
}

@Article{Beg2025Traceable,
  author  = {Beg, Arshad and O'Donoghue, Diarmuid and Monahan, Rosemary},
  journal = {Journal of Software Testing, Verification and Reliability},
  title   = {Traceable and Verifiable Software Requirements: A Synthesis of AI-Enabled Formal Methods},
  year    = {2025},
  note    = {Submitted; Version v1 available on Zenodo},
  doi     = {10.5281/zenodo.17772835},
  url     = {https://doi.org/10.5281/zenodo.17772835},
}

@InProceedings{VeCoGen-Formalise-ICSE2025,
  author    = {Merlijn Sevenhuijsen and Khashayar Etemadi and Mattias Nyberg},
  booktitle = {13th {IEEE/ACM} International Conference on Formal Methods in Software Engineering, FormaliSE@ICSE 2025, Ottawa, ON, Canada, April 27-28, 2025},
  title     = {VeCoGen: Automating Generation of Formally Verified {C} Code With Large Language Models},
  year      = {2025},
  pages     = {101--112},
  publisher = {{IEEE}},
  bibsource = {dblp computer science bibliography, https://dblp.org},
  doi       = {10.1109/FORMALISE66629.2025.00017},
  url       = {https://doi.org/10.1109/FormaliSE66629.2025.00017},
}

@InProceedings{CASPdataset-AISoLA2025,
  author    = {Niclas Hertzberg and Merlijn Sevenhuijsen and Liv K{\aa}reborn and Anna Lokrantz},
  booktitle = {Bridging the Gap Between {AI} and Reality - Third International Conference on Bridging the Gap between {AI} and Reality, AISoLA 2025, Rhodes, Greece, November 1-5, 2025, Selected Papers},
  title     = {{CASP:} An Evaluation Dataset for Formal Verification of {C} Code},
  year      = {2025},
  editor    = {Bernhard Steffen},
  pages     = {63--82},
  publisher = {Springer},
  series    = {Lecture Notes in Computer Science},
  volume    = {16220},
  bibsource = {dblp computer science bibliography, https://dblp.org},
  doi       = {10.1007/978-3-032-07132-3\_5},
  url       = {https://doi.org/10.1007/978-3-032-07132-3\_5},
}

\end{document}